\documentclass[%
 aip,
 jmp,%
 amsmath,amssymb,
 reprint,%
]{revtex4-1}

\usepackage{graphicx}
\usepackage{dcolumn}
\usepackage{bm}
\usepackage{mathrsfs}
\usepackage{amsmath}
\usepackage{epstopdf}
\usepackage{subfigure}
\usepackage{amsfonts}
\usepackage{color}
\usepackage{amssymb}

\begin{document}
\newtheorem{theorem}{Theorem}

\preprint{AIP/123-QED}

\title{Curl flux, coherence and population landscape of molecular systems: nonequilibrium quantum steady state, energy (charge) transport and thermodynamics}

\author{Z.\ D.\ Zhang}
\affiliation{Department of Physics and Astronomy, SUNY Stony Brook, NY 11794, USA}
\author{J.\ Wang}%
\email{jin.wang.1@stonybrook.edu}
\affiliation{Department of Physics and Astronomy, SUNY Stony Brook, NY 11794, USA}
\affiliation{
Department of Chemistry, SUNY Stony Brook, NY 11794, USA
}%
\affiliation{State Key Laboratory of Electroanalytical Chemistry, Changchun Institute of
Applied Chemistry, Chinese Academy of Sciences, Changchun, Jilin, 130022,
P. R. China}

\date{\today}

\begin{abstract}
We established a theoretical framework in terms of the curl flux, population landscape and coherence for non-equilibrium quantum systems at steady state, through exploring the energy and charge transport in molecular processes. The curl quantum flux plays the key role in determining transport properties and the system reaches equilibrium when flux vanishes. 
The novel curl quantum flux reflects the degree of non-
equilibriumness and the time-irreversibility. We found an analytical expression for the quantum flux and its relationship to the environmental pumping (non-equilibriumness quantified by the voltage away from the equilibrium) and the quantum tunnelling. Furthermore, we investigated another quantum signature, the coherence, quantitatively measured by the non-zero off diagonal element of the density matrix. Populations of states give the probabilities of individual states and therefore quantify the population landscape. Both curl flux and coherence depend on steady state population landscape. Besides the environment-assistance which can give dramatic enhancement of coherence and quantum flux with high voltage at a fixed tunnelling strength, the quantum flux is promoted by the coherence in the regime of small tunnelling while reduced by the coherence in the regime of large tunneling, due to the non-monotonic relationship between the coherence and tunneling. This is in contrast to the previously found linear relationship. For the systems coupled to bosonic (photonic and phononic) reservoirs the flux is significantly promoted at large voltage while for fermionic (electronic) reservoirs the flux reaches a saturation after a significant enhancement at large voltage due to the Pauli exclusion principle.  In view of the system as a quantum heat engine, we studied the non-equilibrium thermodynamics and 
established the analytical connections of curl quantum flux to the transport quantities such as energy (charge) transfer efficiency (ETE or CTE), chemical reaction efficiency (CRE), energy dissipation, heat and electric currents observed in the experiments. We observed a perfect transfer efficiency in chemical reactions at high voltage (chemical potential difference). Our theoretical predicted behavior of the electric current with respect to the voltage is in good agreements with the recent experiments on electron transfer in single molecules.
\end{abstract}

\pacs{Valid PACS appear here}
\maketitle

%

\section{Introduction}

Quantum transport is a non-equilibrium phenomena and important in the study of physics, chemistry and biology \cite{Kim05,Laughlin83,Zhang06,Hinds05,Freed72,Kuznetsov99}. Recently energy and electron transports in nanoclusters, quantum dots, single molecules, light harvesting complex and photosynthetic reaction center have been explored in experiments \cite{Tour97,McEuen00,Umansky05,Mate88,Ho04,Ho02}. However, uncovering the underlying mechanisms and global principles of quantum dynamics is still of a challenge, often due to the poor understanding of non-equilibrium nature of the problems and its interplay with the quantum coherence.

So far various theoretical approaches have been used in the study of quantum transport, such as momentum balance equation in the mesoscopic systems \cite{Schoenmaker02}, the fluctuation-dissipation Kubo formula \cite{Kubo57,Prelovsek05} and non-equilibrium Green's function method \cite{Saintjam71,Combesco71,Car08}.
In the approach of Kubo formula, the environment is not explicit and the whole system is in intrinsic equilibrium at long times. Therefore, the current as well as response functions have been calculated as a linear response to an applied external field \cite{Kubo57,Mori58}. 
The non-equilibrium Green's function approach is essentially perturbative with no explicit reservoir, based on the individual equilibrium states of each subsystem at the beginning and then by including the corrections from leads order by order, with respect to coupling between system and leads \cite{Baym89,Smith86}. 
These indicate that these two approaches in principle cannot be effectively applied to the general non-equilibrium steady state, which plays significant role in the quantum transport.

The quantum master equation (QME) is an alternative tool for studying the irreversible dynamics of quantum systems coupled to environments \cite{Breuer02,Spohn80,Haake73}, instead of the perturbative corrections with respect to the equilibrium state. 
 Even at the classical level, the classical master equation (CME) and Fokker-Plank diffusion equation approaches had been already successfully applied to energy transport induced by adenosine triphosphate (ATP) hydrolysis in biochemical systems \cite{Wang12,Qian05}. In particular, a potential and flux landscape theory for non-equilibrium system were developed for studying the global stability and function of the non-equilibrium systems \cite{Wang08,Wang11}. The non-equilibrium systems can be globally quantified by the steady state probability landscape (or population landscape). 
The curl flux is a quantitative measure of the detailed balance breaking, representing the degree of the non-equilibriumness away from the equilibrium.  The relationship between the non-equilibrium landscape and flux to energy pump as well as phase coherence was also studied in detail \cite{Wang12}. For quantum systems, the coherence, as a quantum signature, contributes to the transport \cite{Ghosh11,Ghosh09} and the degree of the non-equilibriumness in addition to the populations \cite{Wolynes92}. This has been confirmed for quantum coherent excitation (charge) transport in light harvesting of photosynthetic reaction center \cite{Fleming07,Xu1992}. 
The quantum transport in terms of many-body description for nanoscale systems and single molecules was also studied \cite{Esposito06,Esposito09}. The unidirectional electron flow in the non-equilibrium ultrafast electron transfer \cite{Zhong12} was observed in recent experiments of photoreduction dynamics of oxidized photolyase \cite{Zhong13}.

In this paper, we will establish a theoretical
framework in terms of curl flux, population landscape and coherence at steady state for the first time for the
non-equilibrium quantum dynamical systems and associated thermodynamics, by exploring the energy transport between different sites in single molecules \cite{Leitner10,Wolynes04} and chemical reaction process (also charge transfer in molecules) \cite{Marcus08,Tao08,Tao09,Ohmine98,Wolynes1988}. The former is coupled to the two heat environments (bosonic) with different temperatures while the latter is coupled to the two chemical environments (fermionic such as electronic baths) with different chemical potentials. The non-equilibriumness of the system can be quantified by a voltage like variable as the difference in temperatures or chemical potentials of the two underlying environments or baths ($T_1-T_2$ or $\mu_1-\mu_2$) away from the equilibrium condition $T_1=T_2$ or $\mu_1=\mu_2$. By starting from the original Hamiltonian coupled with two reservoirs, we derived the QME and from which uncovered the curl flux for non-equilibrium quantum systems at steady state. We found that the curl flux controls the detailed balance breaking and time-irreversibility, providing a measure on the degree of non-equilibriumness and it also plays a key role in determining quantum transport. Both non-equilibriumness and quantum tunneling determine the quantum flux. 
We found analytical expressions for the quantum flux and its relationship to the environmental pumping and quantum tunneling. Furthermore, we investigated another quantum signature, the coherence 
and found the non-monotonic relationship between the steady-state quantum coherence and the tunneling at fixed voltage. Consequently it shows the nontrivial contribution (nonlinear and non-monotonic) of the coherence to the curl flux and quantum transport, in contrast to the previous linear relationship \cite{Manzano12}. Populations of states give the probabilities of individual states and therefore quantify the population landscape. Both curl flux and coherence depend on the steady state population landscape. On the other hand, we uncover the environmental effect, governed by voltage, on curl flux and quantum transport. We found that the environment-assistance can give dramatic enhancement of coherence and quantum flux at high voltage. In view of the system as a quantum heat engine, we studied the non-equilibrium thermodynamics as well as the transport properties such as energy (charge) transfer efficiency (ETE or CTE), chemical reaction efficiency (CRE), energy dissipation and currents observed in the experiments. We established the analytical connection of those dynamical quantities to the non-equilibrium curl quantum flux. The perfect transfer efficiency in chemical reactions was observed at high voltage (chemical potential difference). We have investigated the heat and chemical (electric) currents. Our theoretical predicted behavior of electric current with respect to voltage is in good agreements with the recent experiments on electron transfer in single molecules \cite{Tao08}. In the last section on dynamical decay of coherence, we observed that the decay in time is often faster when the voltage or chemical potential difference measuring the non-equilibriumness away from equilibrium increases.

\section{General theoretical framework for non-equilibrium quantum dynamics in terms of flux, coherence and population landscape}
In this section we will establish a general framework
and formal description on the quantum non-equilibrium
steady state, the curl flux for quantifying the degree of
non-equilibriumness as well as the quantum transport. From next section more details of such theoretical framework will be illustrated by investigating of the energy (charge) transport and thermodynamics of molecular systems. The general Hamiltonian of a quantum system interacting with $M$ environments is of the form
\begin{equation}
\begin{split}
& H_0=\sum_{n,m}H_{nm}|\psi_n\rangle\langle\psi_m|+\sum_{i=1}^M\sum_{\textbf{k},\sigma}\hbar\omega_{\textbf{k}\sigma}a_{\textbf{k}\sigma}^{(i)\dagger}a_{\textbf{k}\sigma}^{(i)}\\[0.2cm]
& H_{int}=\sum_{i,\langle n,m\rangle}\sum_{\textbf{k},\sigma}g_{\textbf{k}\sigma}^{nm(i)}\left(|\psi_n\rangle\langle\psi_m|a_{\textbf{k}\sigma}^{(i)\dagger}+|\psi_m\rangle\langle\psi_n|a_{\textbf{k}\sigma}^{(i)}\right)
\end{split}
\label{01}
\end{equation}
where $\langle n,m\rangle$ indicates that only the pairs of states $n,m$ with energies $E_n<E_m$ are considered. The first term of $H_0$ represents the Hamiltonian of the system and the second term of $H_0$ represents the Hamiltonian of the environments, while the term $H_{int}$  represents the couplings or interactions between the system and environments.  Under the assumption that the environments are of much larger size than the system, we can study the dynamics of the system by tracing out the environments (or reservoirs), which leads to the master equation of the reduced density matrix:

\begin{equation}
\begin{split}
\frac{\partial\rho_S}{\partial t}=\frac{i}{\hbar}\left[\rho_S,H_S\right]- & \frac{1}{2\hbar^2}\sum_{\omega_{\mu}}\sum_{\omega_{\nu}}\gamma_{\mu\nu}(\omega_{\mu})\bigg(A(\omega_{\mu})^{\dagger}A(\omega_{\nu})\rho_S\\[0.2cm]
& -A(\omega_{\nu})^{\dagger}\rho_S A(\omega_{\mu})\bigg)+h.c.+{\cal O}(g^2)
\end{split}
\label{02}
\end{equation}
where the density matrix can be expanded in terms of the coupling strength between system and environments: $\rho(t)=\rho_S(t)\otimes\rho_R(0)+\rho_c(t)$. In the regime of weak coupling, the master equation above can be truncated to the second order and it can be written in Liouville space in which the density matrix forms a column or vector $|\dot{\rho}_S\rangle=\hat{\mathcal{M}}|\rho_S\rangle$. For convenience, we will write the matrix $\mathcal{M}$ as block form, by separating the population (diagonal elements) and coherence terms (off-diagonal elements) of density matrix
\begin{equation}
\begin{split}
 \begin{pmatrix}
   \dot{\rho}_{\textup{p}}\\[0.15cm]
   \dot{\rho}_{\textup{c}}\\
 \end{pmatrix}=\begin{pmatrix}
                \mathcal{M}_{\textup{p}} & \mathcal{M}_{\textup{pc}}\\[0.15cm]
                \mathcal{M}_{\textup{cp}} & \mathcal{M}_{\textup{c}}\\
               \end{pmatrix}
               \begin{pmatrix}
                \rho_{\textup{p}}\\[0.15cm]
                \rho_{\textup{c}}\\
               \end{pmatrix}
\end{split}
\label{03}
\end{equation}
Here, $\mathcal{M}_{\textup{p}}$ represents the transition matrix in population space. $\mathcal{M}_{\textup{c}}$ represents the transition matrix in coherence space (non-off diagonal elements of density matrix). $\mathcal{M}_{\textup{pc}}$ and  $\mathcal{M}_{\textup{cp}}$ represent the coupling transition matrix between population and coherence space.

To study the dynamics of populations we apply the Laplace transform to the coherence components and then
\begin{equation}
\begin{split}
\bar{\rho}_{\textup{c}}(s)=(s-\mathcal{M}_{\textup{c}})^{-1}\mathcal{M}_{\textup{cp}}\bar{\rho}_{\textup{p}}(s)+(s-\mathcal{M}_{\textup{c}})^{-1}\rho_{\textup{c}}(0)
\end{split}
\label{04}
\end{equation}
the inverse Laplace transform to which gives
\begin{equation}
\begin{split}
\rho_{\textup{c}}(t)=\int_0^t e^{\mathcal{M}_{\textup{c}}(t-\tau)}\mathcal{M}_{\textup{cp}}\rho_{\textup{p}}(\tau)d\tau+e^{\mathcal{M}_{\textup{c}}t}\rho_{\textup{c}}(0)
\end{split}
\label{05}
\end{equation}
By substituting Eq.(\ref{05}) into the dynamical equation for population part in Eq.(\ref{03}) we obtain the reduced QME in population space
\begin{equation}
\begin{split}
\partial_t\rho_{\textup{p}}=\mathcal{M}_{\textup{p}}\rho_{\textup{p}}+\int_0^t & \left[\mathcal{M}_{\textup{pc}}e^{\mathcal{M}_{\textup{c}}(t-\tau)}\mathcal{M}_{\textup{cp}}\right]\rho_{\textup{p}}(\tau)d\tau\\[0.2cm]
& +\mathcal{M}_{\textup{pc}}e^{\mathcal{M}_{\textup{c}}t}\rho_{\textup{c}}(0)
\end{split}
\label{06}
\end{equation}
which indicates that the quantum dynamics leads to a memory effect, that is independent of the random collision. Thus the dynamical equations of quantum systems in population space follow integral-differential equations. This significantly increases the complexity for solving them, even on the numerical level. For the quantum steady state at long times, however, a simple form of these equations can be derived, by exactly evaluating the integrals in time domain
\begin{equation}
\begin{split}
\lim_{t\to\infty}\int_0^t e^{\mathcal{M}_{\textup{c}}(t-\tau)}d\tau=-\mathcal{M}_{\textup{c}}^{-1}
\end{split}
\label{07}
\end{equation}
which leads to the reduced QME at steady state
\begin{equation}
\begin{split}
\left(\mathcal{M}_{\textup{p}}-\mathcal{M}_{\textup{pc}}\mathcal{M}_{\textup{c}}^{-1}\mathcal{M}_{\textup{cp}}\right)\rho_{\textup{p}}^{ss}=0
\end{split}
\label{08}
\end{equation}
Notice that the condition of negativity of the eigenvalues of matrix $\mathcal{M}_{\textup{c}}$ is essential, in order to ensure the convergence of the limit in Eq.(\ref{07}). Therefore we can define the transfer matrix as $T_{mn}=\mathcal{A}_{nn,mm}^{\textup{p}}\rho_{mm}^{\textup{p}}$ for $m\neq n$ where $\mathcal{A}_{\textup{p}}\equiv\mathcal{M}_{\textup{p}}-\mathcal{M}_{\textup{pc}}\mathcal{M}_{\textup{c}}^{-1}\mathcal{M}_{\textup{cp}}$ . For $m=n$, $T_{mn}=0$. The transfer matrix can be decomposed into the symmetric and anti symmetric part. $T_{mn} = \frac{T_{mn} + T_{nm}}{2} + \frac{T_{mn} - T_{nm}}{2}$. We can see immediately that the transition matrix $\mathcal{A}_{nn,mm}^{\textup{p},S}=\frac{T_{mn} + T_{nm}}{2}/\rho_{mm}^{\textup{p}}$ corresponding to the first symmetric part of the transfer matrix satisfy the detailed balance condition $(\frac{T_{mn} + T_{nm}}{2}/\rho_{mm}^{\textup{p}} ) \rho_{mm}^{\textup{p}} - (\frac{T_{nm} + T_{mn}}{2}/\rho_{nn}^{\textup{p}}) \rho_{nn}^{\textup{p}} = 0 $. We also can see that the transition matrix $\mathcal{A}_{nn,mm}^{\textup{p},A}=\frac{T_{mn} - T_{nm}}{2}/\rho_{mm}^{\textup{p}}$ corresponding to the second anti-symmetric part of the transfer matrix does not satisfy the detailed balance condition $(\frac{T_{mn} - T_{nm}}{2}/\rho_{mm}^{\textup{p}}) \rho_{mm}^{\textup{p}} - (\frac{T_{nm} - T_{mn}}{2}/\rho_{nn}^{\textup{p}} ) \rho_{nn}^{\textup{p}} \neq 0 $, giving a non-zero steady state flux. Since the transition matrix controls the dynamics of the quantum system, we can see the quantum dynamics can be decomposed into two driving forces. One driving force satisfies the detailed balance condition $\mathcal{A}_{nn,mm}^{\textup{p},S}=\frac{T_{mn} + T_{nm}}{2}/\rho_{mm}^{\textup{p}}$, determined by the steady state distribution, giving the equilibrium part of the contribution to the dynamics. The other force  $\mathcal{A}_{nn,mm}^{\textup{p},A}=\frac{T_{mn} - T_{nm}}{2}/\rho_{mm}^{\textup{p}}$ does not satisfy the detailed balance, giving the flux component of the driving force for the dynamics. This is similar to the classical case \cite{Qian04, Zia, Wang08} in both discrete and continuous case. The detailed balance part of the driving force is largely dependent on the diagonal element of the density matrix: the steady state population. Populations of states give the probabilities of individual states and therefore quantify the population landscape. Both curl flux and coherence depend on steady state population landscape.
We can investigate further the properties of the non-zero flux which breaks the detailed balance. The quantum flux has curl nature and  can be decomposed further into a sum of fluxes of various loops.

By introducing ${}_{\alpha}\mathcal{A}_{nm}^p\rho_m^p=\textup{min}\left(\mathcal{A}_{nm}\rho_m^p,\mathcal{A}_{mn}\rho_n^p\right)$, the total transition rate matrix that describes the non-equilibrium quantum flux between different pairs of states, can be defined as follows
\begin{equation}
\begin{split}
c_{mn}=\mathcal{A}_{nm}^p\rho_m^p-{}_{\alpha}\mathcal{A}_{nm}^p\rho_m^p
\end{split}
\end{equation}
which can be decomposed into sum of the fluxes of various closed loops if the following theorem is satisfied

\begin{theorem}
The transition rate matrix $c$ can be decomposed to $c=\sum_{i=1}^Q R^{(i)}$ where $R^{(i)}$ is the $i$-th closed curl matrix (circular and divergent free), if the following conditions are satisfied:
\begin{description}
\item{\rm (1).} $c_{mn}\geq 0$ for $m\ne n$ and $c_{nn}=0$;

\item{\rm (2).} $c_{mn}c_{nm}=0$ for $m\neq n$;

\item{\rm (3).} $\sum\limits_{m}c_{mn}=\sum\limits_{n}c_{mn}$
\end{description}

\end{theorem}
which can be understood as follow: The condition (2) in theorem actually means the unidirection of the flux (e.g.: $c_{12}c_{21}=0$ indicates only one direction of the flow can survive !), based on the requirement in condition (1); condition (3) is equivalent to the stationary distribution of population at steady state and the conservation of total population as well. The above flux-decomposition theorem was mathematically proven \cite{Qian04} at the classical level. Here since we reduce the quantum master equation in the Markov chain form at steady state, we can apply the decomposition to investigate the non-equilibrium properties of quantum flux and also quantum transport at steady state, which will be shown in details in subsequent sections. We should notice that although the Markov chain looks similar as the classical case, the coefficients or the rates $\mathcal{A}_{nn,mm}$ and populations $\rho_m$ (diagonal element of the density matrix) are significantly influenced by the quantum coherence (off-diagonal elements of the density matrix) through the dimensional reduction (integrating out the coherence term and obtain the master equation for reduced density matrix in population space). In summary, for non-equilibrium quantum steady state, we can decompose the rate dynamics into the detailed balance preserving part (mainly determined by the population landscape), and detailed balance breaking flux part, which can be further decomposed into the sums of the fluxes of the loops. Both parts depend on quantum coherence. We will focus our attention of this study on the curl flux and quantum coherence. We will discuss further in details on the non-equilibrium quantum landscape in the forthcoming studies. 

\section{Hamiltonian and Quantum Master Equation}
We will illustrate our ideas in the examples of the quantum energy and charge transfer processes. 

\subsection{Coupled to heat (bosonic) reservoirs}
Energy transfer in molecules happens between donor and acceptor sites, after being excited from ground state. To be simple, we assume the excitation energies of these two sites are of a small difference, namely, $\varepsilon_2-\varepsilon_1\ll \textup{min}(\varepsilon_1,\varepsilon_2)$. In the language of excitons, this system can be modelled as asymmetric double wells, as shown schematically in Fig.1. Since we would discuss the transport between different sites in molecules, it is clearer to describe the system in local representation. 
The subspace relevant to our discussion is spanned by the following two types of excitation in addition to the ground state.
\begin{equation}
|\Omega\rangle=C_g^{\dagger}|0\rangle,\quad |1\rangle=C_1^{\dagger}C_g|\Omega\rangle,\quad |2\rangle=C_2^{\dagger}C_g|\Omega\rangle
\label{1}
\end{equation}
where $|0\rangle$ stands for the vacuum, $C_i$ and $C_j^{\dagger}$ are the annihilation and creation operators for electrons in molecules. The transport in molecules can be modelled by the system interacting with two identical reservoirs with different temperatures. The free and interaction Hamiltonian then read
\begin{equation}
\begin{split}
& H_S=E_g|\Omega\rangle\langle\Omega|+\varepsilon_1\eta_1^{\dagger}\eta_1+\varepsilon_2\eta_2^{\dagger}\eta_2+\Delta(\eta_1^{\dagger}\eta_2+\eta_2^{\dagger}\eta_1)\\[0.2cm]
& \quad\quad\quad\quad H_R=\sum_{\textbf{k},p}\hbar\omega_{\textbf{k}p}a_{\textbf{k}p}^{\dagger}a_{\textbf{k}p}+\sum_{\textbf{q},s}\hbar\omega_{\textbf{q}s}b_{\textbf{q}s}^{\dagger}b_{\textbf{q}s}
\end{split}
\label{2}
\end{equation}
\begin{equation}
H_{int}=\sum_{\textbf{k},p}\lambda_{\textbf{k}p}\left(\eta_2^{\dagger}a_{\textbf{k}p}+\eta_2a_{\textbf{k}p}^{\dagger}\right)+\sum_{\textbf{q},s}\lambda_{\textbf{q}s}\left(\eta_1^{\dagger}b_{\textbf{q}s}+\eta_1 b_{\textbf{q}s}^{\dagger}\right)
\label{4}
\end{equation}
where $\eta$ and $\eta^{\dagger}$ are the annihilation and creation operators for excitons 
which obey the Fermi-Dirac statistics: $\{\eta_a,\eta_b^{\dagger}\}=\delta_{ab}$, $\{\eta_a,\eta_b\}=0$. The scattering between excitons is neglected here. 
The annihilation and creation operators $a(b)$ and $a^{\dagger}(b^{\dagger})$ for the environments (reservoirs) satisfy Bose-Einstein relations: $[a_{\textbf{k}p},a_{\textbf{k}'p'}^{\dagger}]=\delta_{\textbf{k}\textbf{k}'}\delta_{pp'}$, $[a_{\textbf{k}p},a_{\textbf{k}'p'}]=0$. $\Delta$ represents the electronic coupling (tunnelling strength) between the two sites. Here we do not include the vibrational degree of freedoms of the nuclei, due to their fast relaxation within the time scale of  $10^{-12}$s, which is much shorter than the time scale of electronic excitation. However, the electric dephasing occurs on a comparable timescale to vibrational relaxation in the light-harvesting and Fenna-Matthews-Olson complexes, therefore we will include this effect in our future work, since this effect goes beyond the scope of current paper. $p$ and $s$ denote the polarizations of the boson (either radiation or phonon) field. 
In Eq.(\ref{4}) the rotating-wave approximation \cite{Landau77} was applied to exciton-photon interaction term due to the dominant contribution by real absorption and emission. Only the single-exciton process is important to energy transport and then in the single-exciton manifold the Hamiltonians (\ref{2}) and (\ref{4}) are taken the forms of $H_S=E_g|\Omega\rangle\langle\Omega|+\varepsilon_1|1\rangle\langle1|+\varepsilon_2|2\rangle\langle2|+\Delta\left(|1\rangle\langle 2|+|2\rangle\langle 1|\right)$ and
\begin{figure}
\centering\includegraphics[width=3.26in,height=1.9in]{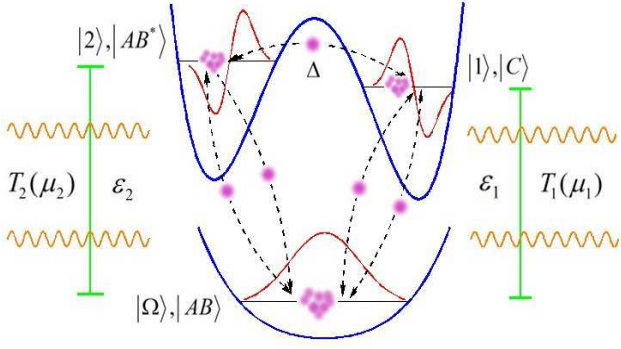}
\caption{(Color online) \textit{Schematic illustration of the Energy Transport in Single Molecules and Chemical Reaction} $AB\leftrightarrow C$, as discussed in details in our paper. The two reservoirs keep their own temperatures or chemical potentials, respectively. The energy or charge (chemical species) will flow from state $|2\rangle$ (the intermediate $|AB^*\rangle$) to state $|1\rangle$ ($|C\rangle$)}
\label{Fig.1}
\end{figure}

\begin{equation}
\begin{split}
H_{int}=\sum_{\textbf{k},p}\lambda_{\textbf{k}p} & \left(\sigma_{2g}^+ a_{\textbf{k}p}+\sigma_{2g}^-a_{\textbf{k}p}^{\dagger}\right)\\
& +\sum_{\textbf{q},s}\lambda_{\textbf{q}s}\left(\sigma_{1g}^+ b_{\textbf{q}s}+\sigma_{1g}^-b_{\textbf{q}s}^{\dagger}\right)
\end{split}
\label{5}
\end{equation}
where the creation and annihilation of excitons were replaced by transition, namely, $\eta_i^{\dagger}\rightarrow\sigma_{ig}^+\equiv |i\rangle\langle \Omega|$ and $\eta_i\rightarrow\sigma_{ig}^-\equiv |\Omega\rangle\langle i|\quad (i=1,2)$, in the single-exciton manifold. The quantum mechanical tunnelling between different sites delocalizes the wave function over the diameter of molecule, which provides an intuitive understanding at first step of effects of quantum coherence on transport.
Using Bogoliubov transformation \cite{Landaustatmech77} the system Hamiltonian in Eq.(\ref{2}) is diagonalized by switching into delocalized representation. Therefore it is convenient for us to do the further derivation for the quantum master equation with the help of the interaction picture \cite{Scully97,Breuer02} (This is because only the interaction term will appear in the total Hamiltonian and the calculation can then be simplified).
Since we are interested in the evolution of the variables associated with the system only, the equation for the reduced density matrix in the subspace need to be obtained, by performing a partial trace over the reservoir freedoms. 
As the coupling strength in {\it Quantum Electrodynamics} is of the order of fine structure constant, the whole solution of density operator can be written as $\rho_{SR}\left(t\right)=\rho_S\left(t\right)\otimes\rho_{R_1}(0)\otimes\rho_{R_2}(0)+\rho_c\left(t\right)$ with the traceless term in a higher order of coupling, and the system is assumed to be memoryless, which is so-called the Markovian approximation. This is valid when the correlation time scale of reservoirs is much shorter than the time scale for the dynamics of the system. In other words, the reservoirs have the white noise feature, which is applicable for the system maintaining in thermal equilibrium. Therefore the master equation for reduced density matrix reads
\begin{equation}
\begin{split}
 \frac{\textup{d}\rho_S}{\textup{d}t}= & \frac{i}{\hbar}\left[\rho_S,H_S\right]-\frac{1}{\hbar^2}e^{-iH_St/\hbar}\textup{Tr}_{R_1R_2}\\[0.2cm]
& \int_0^t \textup{d}s\left[\tilde{H}_{int}(s),\left[\tilde{H}_{int}(t),\tilde{\rho}_S(t)\otimes\rho_{R}(0)\right]\right]e^{iH_St/\hbar}
\label{11}
\end{split}
\end{equation}
On inserting the energy in the interaction picture into the equation of motion Eq.(\ref{11}), then tracing out the environments, the QME in our model is arrived, in the localized representation
\begin{equation}
\begin{split}
\dot{\rho} _S=\frac{i}{\hbar}\left[\rho_S,H_S\right]-\frac{1}{2\hbar^2}\mathcal{D}\left(\rho_S\right)
\end{split}
\label{12}
\end{equation}
where the form of super operator $\mathcal{D}$ will be given in detail in appendix A. $n_{\omega}=\left[\textup{exp}\left(\frac{\hbar\omega}{k_B T}\right)-1\right]^{-1}$ is the Bose average occupation on frequency $\omega$ at temperature $T$. The Weisskopf-Wigner approximation that the upper limit of integral over time can be extended to infinity due to the rapid oscillation of the integrand for $s\ll t$ was used in deriving Eq.(\ref{12}). 
Hence the decay rates induced by photon reservoirs are $\Gamma_a/\hbar^2=\frac{V}{4\pi^2\hbar^2}\int\textup{d}^3\textbf{k}\ \lambda_{\textbf{k}}^2\delta\left(\omega_{\textbf{k}}-\omega'_{ag}\right)=\frac{4\pi^2\lambda}{\hbar}\frac{{\omega'}_{ag}^3r_m^3}{8\pi^3c^3},a=1,2$ 
 where $r_m\sim 30$nm is the separation between the complex molecules, such as mesobiliverdin (MBV) $\&$ dihydrobiliverdin (DBV) molecules in light harvesting complex \cite{collini10}, the spectral density $J(\omega)=4\pi^2\hbar\lambda\frac{\omega^3r_m^3}{8\pi^3c^3}$ and $\lambda$ is the reorganization energy. After some mathematical procedures, we can derive the compact form of QME in Liouville space: $\partial_t |\rho\rangle=\mathcal{M}|\rho\rangle$, by writing the density matrix as a super-vector: $|\rho\rangle=\left(\rho_{gg},\rho_{11},\rho_{22},\rho_{12},\rho_{21}\right)^{\textup{T}}$ where $\mathcal{M}_{22}^{11}=\mathcal{M}_{11}^{22}=0$, $\mathcal{M}_{21}^{12}=\mathcal{M}_{12}^{21}=0$.
Matrix $\mathcal{M}$ is determined by Eq.(\ref{12}). The analytical expressions for the elements in $\mathcal{M}$ will be given in Appendix A. It is easy to verify that $\sum_{a=g}^2\mathcal{M}_{kl}^{aa}=0$ which reveals the charge conservation. Moreover, the coherence terms quantified by the non-zero off diagonal elements of the density matrix $\rho_{g1}$ and $\rho_{g2}$ as well as their complex conjugates are absent from QME in that they are only entangled to themselves in the equations of dynamical evolution. Therefore only the coherence between excitations ($\rho_{12}$ and $\rho_{21}$) contributes to our conclusions and the discussion can be restricted into the 5-dimensional space.

\subsection{Coupled to chemical (fermionic) reservoirs}
To describe the transport of chemical recombination dissociation reaction $AB\leftrightarrow C$ with the vibrationally excited intermediate $AB^*$, we model the quantum system interacting with two chemical reservoirs or leads with different chemical potentials which provide the effective chemical pumping for the system by collisions \cite{Marcus08}. The three quantum states are denoted by $|AB\rangle$, $|AB^*\rangle$ and $|C\rangle$ as schematically shown in Fig.1. The Hamiltonian is of the similar form as the one with bosonic reservoirs:
\begin{equation}
\begin{split}
& H_S=E_g|\Omega\rangle\langle\Omega|+\varepsilon_1 c_1^{\dagger}c_1+\varepsilon_2 c_2^{\dagger}c_2+\Delta(c_1^{\dagger}c_2+c_2^{\dagger}c_1)\\[0.2cm]
& \quad\quad\quad\quad H_R=\sum_{\textbf{k}}\hbar\nu_{\textbf{k}}a_{\textbf{k}}^{\dagger}a_{\textbf{k}}+\sum_{\textbf{q}}\hbar\nu_{\textbf{q}}b_{\textbf{q}}^{\dagger}b_{\textbf{q}}\\[0.2cm]
& H_{int}=\sum_{\textbf{k}}f_{\textbf{k}}\left(c_2^{\dagger}a_{\textbf{k}}+c_2a_{\textbf{k}}^{\dagger}\right)+\sum_{\textbf{q}}f_{\textbf{q}}\left(c_1^{\dagger}b_{\textbf{q}}+c_1b_{\textbf{q}}^{\dagger}\right)
\end{split}
\end{equation}
where the operators for chemical reservoirs obey the Fermi-Dirac statistics: $\{a_{\textbf{k}},a_{\textbf{k}'}^{\dagger}\}=\delta_{\textbf{k}\textbf{k}'}$ and $\{b_{\textbf{q}},b_{\textbf{q}'}^{\dagger}\}=\delta_{\textbf{q}\textbf{q}'}$. $\Delta$ describes the conversion between states $|AB^*\rangle$ and $|C\rangle$ by the tunnelling through the barrier. The rotating-wave approximation was applied as well and the occupation will be replaced by fermionic type: $n_{\omega}^{\mu}=\left[\textup{exp}\left(\frac{\hbar\omega-\mu}{k_B T}\right)+1\right]^{-1}$. Instead of the linear dependence on wave vector in radiation fields, the dispersion relation in solvent or semiconductor lead can be approximated by a parabolic law, namely, $\varepsilon_{\textbf{k}}\simeq \frac{\hbar^2k^2}{2m^*}$ where $m^*$ is the effective mass. Therefore the decay rate reads $\Gamma_a/\hbar^2=\frac{1}{\hbar^2}\int\textup{d}\nu D(\nu)f_{\nu}^2\delta(\nu-\omega'_{ag})$ where $D(\nu)\sim \sqrt{\nu}$ is the density of states, and the spectrum density is $J(\nu)=D(\nu)f_{\nu}^2=\frac{\hbar\lambda}{2\pi}\varphi(\nu)$, where $\varphi(\nu)$ is a smooth and dimensionless function, with the magnitude on the order of $\sim 1$. Hence the remaining procedures are the same as the bosonic reservoir case above and we will skip the details to avoid redundancy. Finally the reduced QME for fermionic baths can be derived: $\partial_t|\rho\rangle=\mathcal{M}|\rho\rangle$ in Liouville space. Based on these preparations we are able to develop the quantum curl flux decomposition which will be shown in next section.

\section{Curl decomposition and non-equilibrium quantum flux}
After deriving the QME in detail in last section, we will, in this section, introduce the curl decomposition for quantum steady state which is crucial because it generates a novel quantum flux for quantifying the quantum transport and the flux directly reflects the detailed balance breaking and time-irreversibility. The similar decomposition for classical open chemical system at steady state was discussed before \cite{Qian06,Qian09}. By eliminating the off-diagonal components in density matrix from the reduced QME through Laplace transform we can map the reduced QME into population space, i.e., $\partial_t|\rho\rangle=\mathcal{H}|\rho\rangle$ where $\mathcal{H}$ is a matrix with integral kernel (shown in Appendix B) in Liouville space which is the extended Hilbert space and density matrix elements form a supervector.
Consequently the quantum effects on transport is somewhat equivalent to the memory effect, which has nothing to do with the mechanisms of collision. This is absent in classical theory described by CME. In this study, we are interested in the quantum non-equilibrium steady state so that we evaluate the integrals by extending the upper limit to $\infty$ to obtain the reduced QME at steady state
\begin{equation}
 \begin{pmatrix}
  \mathcal{M}_{gg}^{gg}-2\mathcal{C}_{gg}^{gg} & \mathcal{M}_{11}^{gg}-2\mathcal{C}_{11}^{gg} & \mathcal{M}_{22}^{gg}-2\mathcal{C}_{22}^{gg}\\[0.25cm]
  \mathcal{M}_{gg}^{11}-2\mathcal{C}_{gg}^{11} & \mathcal{M}_{11}^{11}-2\mathcal{C}_{11}^{11} & -2\mathcal{C}_{22}^{11}\\[0.25cm]
  \mathcal{M}_{gg}^{22}-2\mathcal{C}_{gg}^{22} & -2\mathcal{C}_{11}^{22} & \mathcal{M}_{22}^{22}-2\mathcal{C}_{22}^{22}\\
 \end{pmatrix}
\begin{pmatrix}
  \rho_{gg}\\[0.25cm]
  \rho_{11}\\[0.25cm]
  \rho_{22}\\
 \end{pmatrix}
 =0
\label{17}
\end{equation}
where $\mathcal{C}_{kl}^{mn}\equiv \textup{Re}\left(\mathcal{M}_{12}^{mn}\mathcal{M}_{kl}^{12}/\mathcal{M}_{12}^{12}\right)$ and $\mathcal{M}$ are defined before. 
As $\mathcal{M}_{12}^{12}$ governs the decay rate in the integral kernel, such memorable effect is significant for large $|\textup{Re}\left(\mathcal{M}_{12}^{12}\right)|$ while it becomes tiny for small $|\textup{Re}\left(\mathcal{M}_{12}^{12}\right)|$. The reduced QME in Eq.(\ref{17}) is of the same form as the CME within Markovian approximation, but with a different explanation: \textit{the quantum effect has already been contained and reflected through $\mathcal{C}$-matrix by the exact evaluation of the integral kernel for memory in our QME}, in contrast to previous work with the additional second Markovian approximation \cite{Cao12,Wu12}. In classical open systems the $\mathcal{C}$-matrix vanishes.

\subsection{Curl quantum flux determined by the non-equilibriumness and tunnelling}
\subsubsection{The analytical forms}$\quad$Next we need to introduce the non-equilibrium quantum flux in order to investigate the quantum transport. First the transfer matrix has to be defined: $T_{kl}^{mn}=\mathcal{A}_{mn}^{kl}\rho_{mn}$ with zero diagonal element. Then this $T$-matrix can be decomposed into the following form
\begin{equation}
T=\begin{pmatrix}
   0 & \mathcal{A}_{gg}^{11}\rho_{gg} & \mathcal{A}_{22}^{gg}\rho_{22}\\
   \mathcal{A}_{gg}^{11}\rho_{gg} & 0 & \mathcal{A}_{11}^{22}\rho_{11}\\
   \mathcal{A}_{22}^{gg}\rho_{22} & \mathcal{A}_{11}^{22}\rho_{11} & 0\\
  \end{pmatrix}
 +\begin{pmatrix}
   0 & 0 & \mathcal{J}_q\\
   \mathcal{J}_q & 0 & 0\\
   0 & \mathcal{J}_q & 0\\
  \end{pmatrix}
\label{18}
\end{equation}
The reduced QME in population space directly gives the expression $\mathcal{J}_q=\mathcal{A}_{22}^{11}\rho_{22}-\mathcal{A}_{11}^{22}\rho_{11}$. In Eq.(\ref{18}) the $1^{\textup{st}}$ term of the transfer matrix describes the equilibrium with detailed balance preserved; The $2^{\textup{nd}}$ term is circular that we call 'non-equilibrium quantum flux', which plays a crucial role in determining the transport properties of open quantum systems, such as entropy production (EPR), dissipation and efficiency. Moreover, the curl flux matrix in Eq.(\ref{18}) is closed at steady state, by the application of Theorem 1 before. 
By solving the QME at steady state under the further approximation $|\Delta|\ll \textup{min}(\varepsilon_1,\varepsilon_2)$, $n_{\varepsilon}\simeq \frac{1}{2}(n_{\omega'_{1g}}+n_{\omega'_{2g}})$ and $\Gamma=\frac{1}{2}(\Gamma_1+\Gamma_2)$, we can obtain the expression for quantum flux in our model
where ($\hbar\omega\equiv\varepsilon_2-\varepsilon_1$) 
\begin{equation}
\begin{split}
\mathcal{J}_q^b=\frac{2\Gamma}{\hbar^2}\frac{v^b\frac{\Delta^2}{\hbar^2\omega^2}}{1+4u^b\frac{\Delta^2}{\hbar^2\omega^2}},\quad \mathcal{J}_q^f=\frac{2\Gamma}{\hbar^2}\frac{v^f\frac{\Delta^2}{\hbar^2\omega^2}}{1+4u^f\frac{\Delta^2}{\hbar^2\omega^2}}
\end{split}
\label{20}
\end{equation}
where the function $v$ provides a measure for the effective voltage and detailed balance breaking induced from environments. {\it The flux describes how much probability flows in a uni-direction from one site to another in unit time}.
The forms of $u$ and $v$ are given in Appendix B. Therefore, the function $v$ quantifies the degree of non-equilibriumness away from the equilibrium. Then from the expressions of quantum flux in Eq.(\ref{20}), we can see the quantum transport quantified by the non-equilibrium quantum flux is determined by two factors: non-equilibriumness quantified by the effective voltage away from equilibrium and the quantum tunnelling. When the effective voltage is zero, the system is at quantum equilibrium with no quantum flux or quantum transport. On the other hand, when the effective voltage increases, the quantum flux increases. The degree of non-equilibriumness drives the quantum transport. In this model, the quantum transport is realized by tunnelling from one site to another. When $\Delta=0$, the flux is zero and there is no quantum transport. The quantum flux increases as the tunneling increases until the tunnelling becomes big and the quantum flux reaches a plateau. The quantum tunnelling promotes the quantum transport at moderate regime, in that there is no effective barrier any more. The further increasing tunnelling will not increase the quantum transport further at very large tunnelling strength. 
See next section for further detailed explanations in a different angle.

\subsubsection{Numerical Results}$\quad$Fig.9(a) and 9(b) in appendix A give the comparison between our analytical formula for quantum flux Eq.(\ref{20}) and the results from numerical simulation, as functions of bias (for bosonic bath it is temperature difference while for fermionic bath it is chemical potential difference). 
Fig.3 shows the variations of quantum flux for bosonic and fermionic reservoirs with respect to voltage as well as tunneling strength. Qualitatively, it is known that the quantum tunneling gives rise to the so-called dark states which is the superposition of two excited states to make the transfer of energy or charge being enhanced, but such contribution will reach the saturation at large value of tunneling. On the other hand it is found that large bias leads to significant enhancement of flux, which indicates that far-from-equilibrium rather than near-to-equilibrium is crucial for the enhancement of quantum flux and the transport. Furthermore, compared to bosonic case, a sharp increase of flux occurs after a particular value of bias, which is about 0.8eV in our plot. This is because of the Fermi-Dirac distribution, where the density of excitations will be much suppressed as the energy becomes larger than fermi energy. Another distinction is that for bosonic baths the increase of flux becomes sharper as the system deviates from equilibrium while for fermionic baths the flux reaches saturation as the system deviates very far from equilibrium. This is due to the Pauli exclusion principle that 
for fermions the occupation for each frequency is no more than one.

\subsection{The nontrivial relationships among coherence, tunneling, non-equilibriumness and quantum flux}

\subsubsection{Non-monotonic relationship between tunneling and coherence}$\quad$Based on QME and the approximation above, we can obtain the quantum coherence. Furthermore the connection of quantum flux to coherence reads
$\mathcal{J}_q^{b(f)}=\frac{2\Delta}{\hbar}\times|\textup{Im}\rho_{12}|$. Notice that the environmental effect was included in the coherence. Right now we can conclude that coherence enhances the non-equilibrium flux through a linear law when fixing the tunneling strength, which reveals the important distinction of the properties of non-equilibrium quantum system from classical description. In order to see how environments and quantum coherence affect the flux as well as ETE (CRE), we can write the coherence in terms of the tunneling strength
\begin{equation}
|\textup{Im}\rho_{12}|=\frac{\Gamma v}{\hbar^2\omega}\frac{\frac{\Delta}{\hbar\omega}}{1+4u\frac{\Delta^2}{\hbar^2\omega^2}}
\label{21}
\end{equation}
As we can see the coherence has a non-trivial non-monotonic dependence on tunneling as shown in Fig.2(a) and moreover the quantum coherence is also promoted by voltage when fixing the tunneling. There is a peak of coherence at $\Delta_c=\frac{\hbar\omega}{2\sqrt{u}}$. Then for large $\Delta$ (which indicates a large coupling and transport between the molecules)
the height of barrier in the middle is effectively lowered. The state $|2\rangle$ is switched to $|2'\rangle$ with excitation energy $\varepsilon'_2\simeq\bar{\varepsilon}+\Delta$ where $\bar{\varepsilon}\equiv\frac{1}{2}(\varepsilon_1+\varepsilon_2)$ and further the distribution of bosons (fermions)$\sim e^{-\beta\Delta}$. From the quantum-classical correspondence we know that $|2'\rangle$ is approaching the classical limit, which means the behavior of particles at this state is close to classical motion. Thus in fact $|2'\rangle$ becomes a quasi-classical state, which leads to the reduction of coherence for large tunneling. 
On the other hand, 
there is an upper limit for tunneling, roughly $\Delta\sim\bar{\varepsilon}$, since the lowering of barrier gives rise to the shallowness of the first well, which as a result, leads to the vanishing of bound state if $\Delta\gg\bar{\varepsilon}$.

\subsubsection{Non-monotonic relationship between coherence, non-equilibriumness and quantum flux}$\quad$We next explore the relationship between coherence and quantum flux. By eliminating $\Delta$ in Eq.(\ref{20}) and (\ref{21}) it leads to the alternative expression of flux
\begin{equation}
\mathcal{J}_q^{b(f)}=\frac{\Gamma v}{4\hbar^2u}\left(1\pm\sqrt{1-\frac{16\hbar^4\omega^2u|\textup{Im}\rho_{12}|^2}{\Gamma^2v^2}}\right)
\label{23}
\end{equation}
where $-$ and $+$ correspond to $0<\Delta<\frac{\hbar\omega}{2\sqrt{u}}$ and $\Delta>\frac{\hbar\omega}{2\sqrt{u}}$, respectively. For the small tunneling the flux has the asymptotic form $\mathcal{J}\simeq\frac{2\hbar^2\omega^2}{\Gamma v}|\textup{Im}\rho_{12}|^2$. The behavior of flux with respect to coherence is plotted in Fig.2(b).
As is shown, $\mathcal{J}$ first monotonically increases by the improvement of coherence, but then the
reduction of the coherence gives rise to the enhancement of the flux. This is mainly because of the non-monotonic behavior of coherence as a function of quantum tunneling discussed above. In the second regime, the increasing tunneling still improves quantum transport but the quasi-classical limit reduces the coherence.

As shown in Eq.(\ref{23}) and Fig.2(c), we see the non-monotonic behavior of the flux with respect to the non-equilibriumness characterized by the  voltage at fixed coherence. 
Again, this non-trivial relationship of flux versus non-equilibriumness quantified by the effective voltage is from the non-trivial relationship between the tunneling and coherence discussed before. Physically Fig.2(c) can be explained as the consumption of the energy is used for keeping the coherence such that at small tunneling $\Delta$ needs to reduce in order to balance out the improvement of coherence by voltage since $\Delta$ enhances coherence. Hence equivalently much energy absorbed from environments is used to fix the coherence but less to improve flux. In contrast, at large tunneling $\Delta$ needs to increase, in order to balance out the improvement of coherence by voltage, therefore much more energy is used to enhance the flux instead of keeping coherence.

\begin{figure}
\centering
$\begin{array}{cc}
 \includegraphics[width=1.6in,height=1.22in]{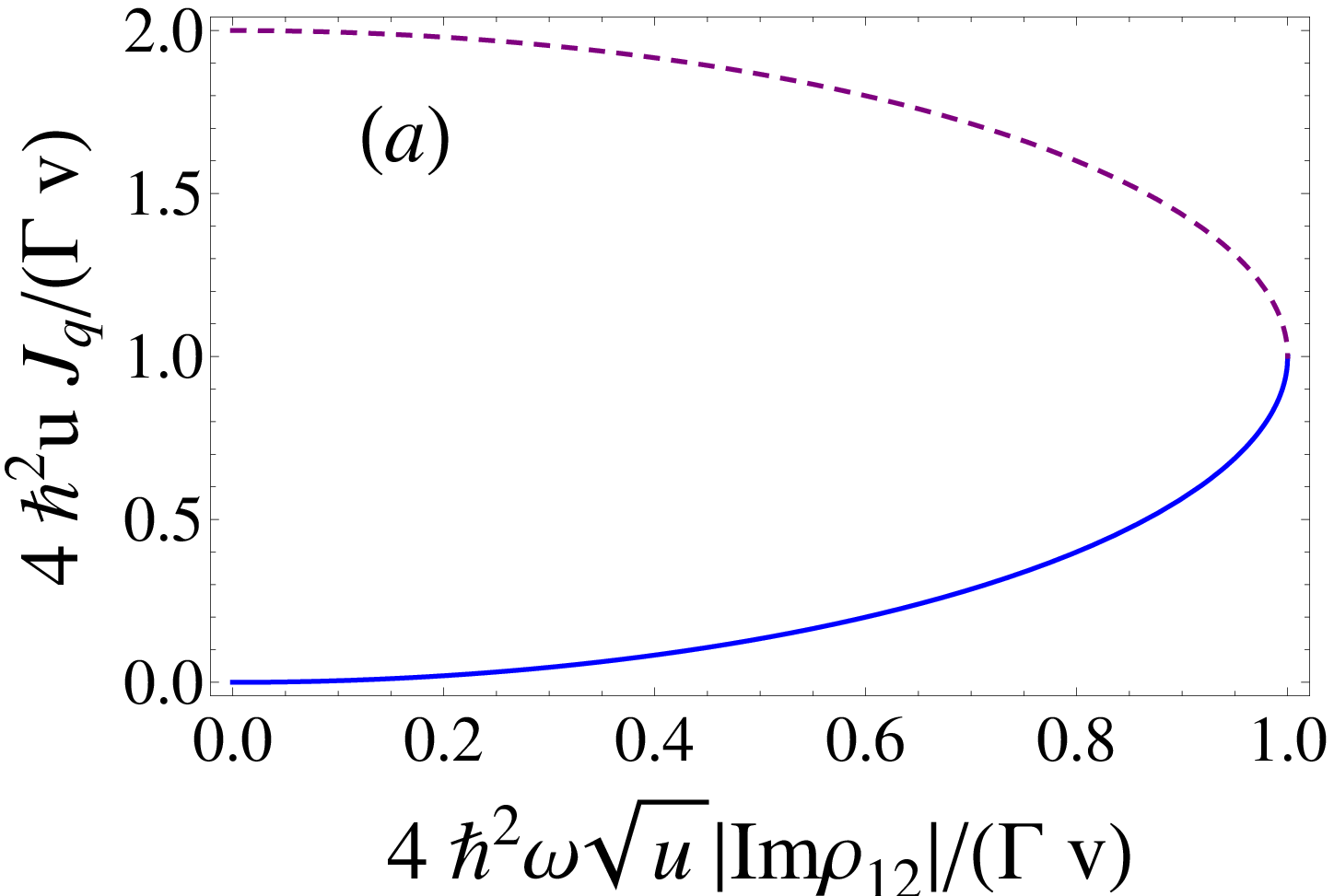}
&\includegraphics[width=1.6in,height=1.22in]{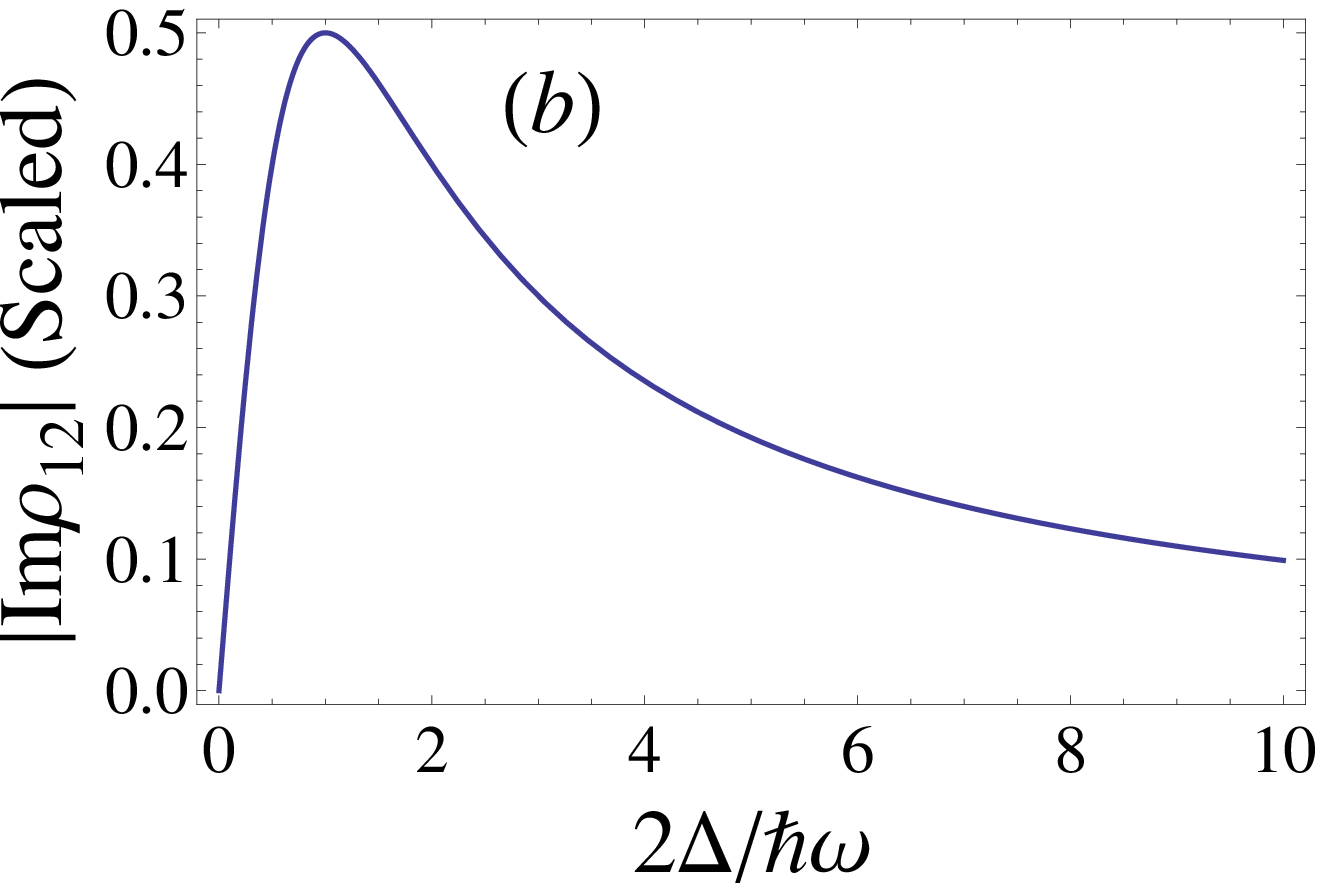}\\
\includegraphics[width=1.6in,height=1.22in]{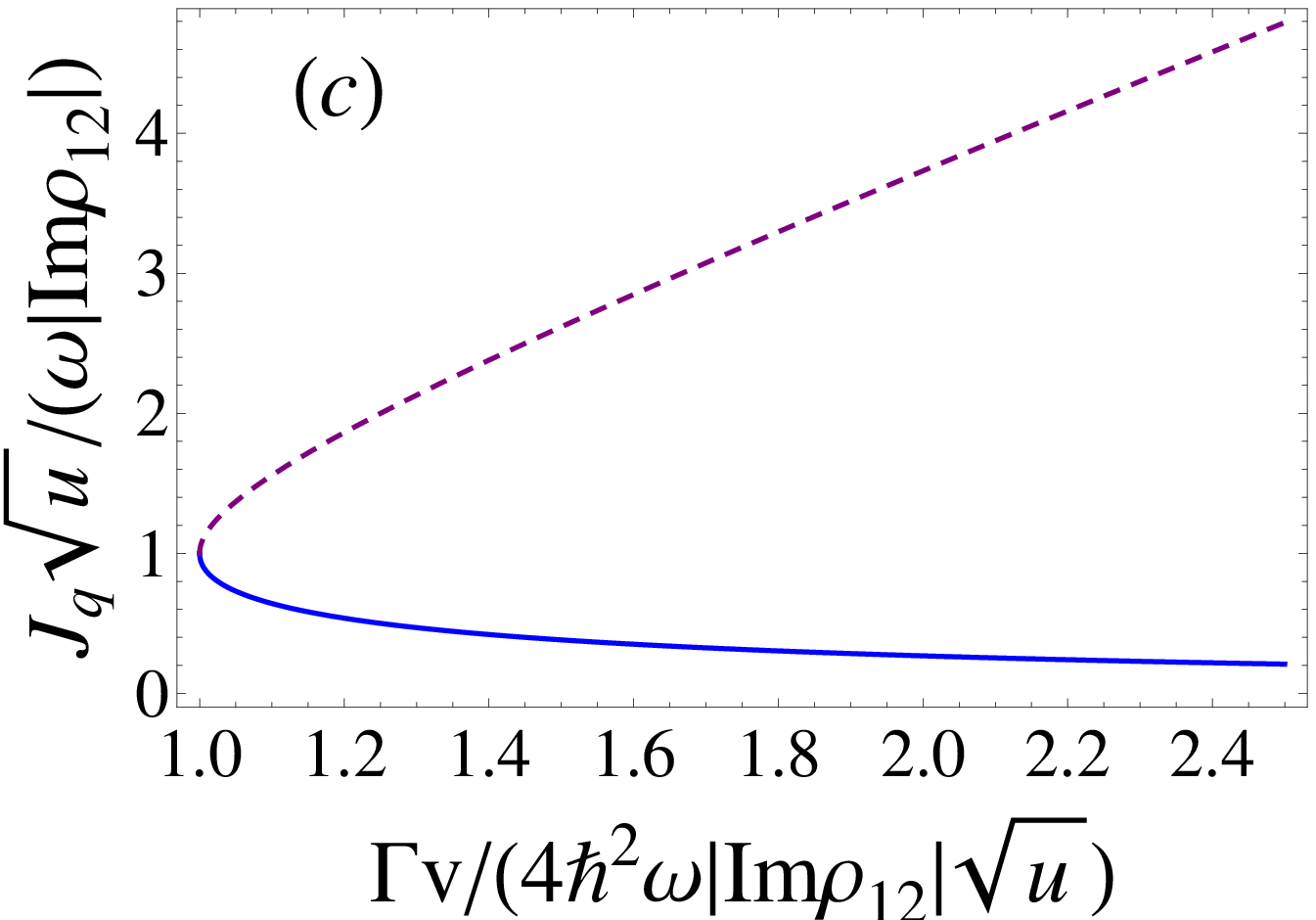}
&\includegraphics[width=1.6in,height=1.19in]{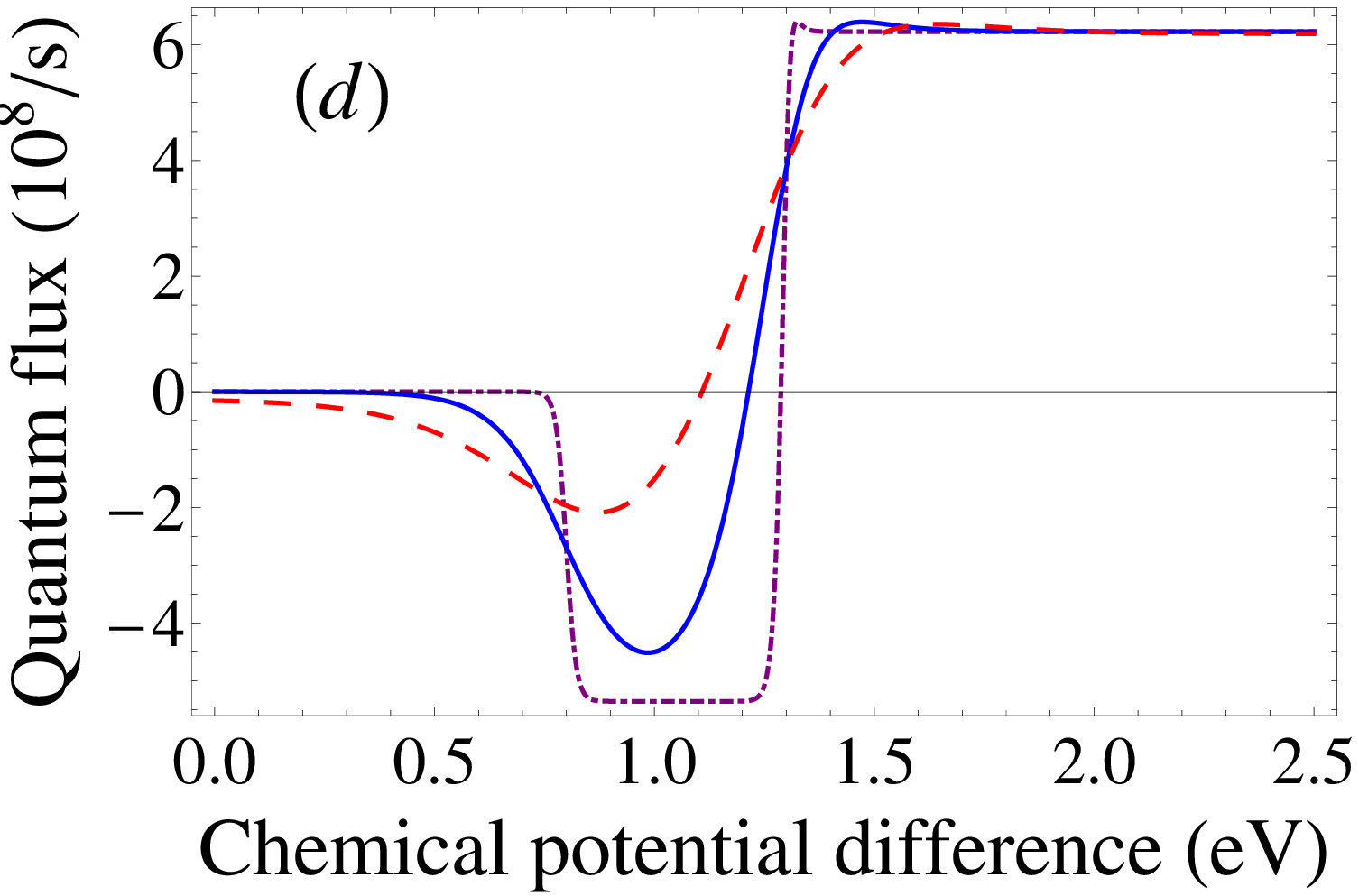}\\
\end{array}$
\caption{(Color online) Quantum flux varies as a function of (a) coherence and (c) voltage function $v$ for both bosonic and fermionic cases. Flux and coherence are scaled by $\frac{\Gamma v}{4\hbar^2 u}$ and $\frac{\Gamma v}{4\hbar^2\omega\sqrt{u}}$, respectively. Blue and red curves correspond to $\Delta<\frac{\hbar\omega}{2\sqrt{u}}$ and $\Delta>\frac{\hbar\omega}{2\sqrt{u}}$, respectively; (b) Imaginary part of quantum coherence as a function of tunneling, represented by $\frac{2\Delta}{\hbar\omega}$; (d) Quantum flux varies as a function of chemical potential difference. Purple, blue and red lines are for $T=130$K, 900K and 1800K, respectively. Standard parameters are $\varepsilon_1=0.9$eV, $\varepsilon_2=1.2$eV, $\Delta=0.2$eV, $\lambda=21$cm$^{-1}$ and $\mu_1=0$}
\label{Fig.2}
\end{figure}


We should note that tunneling and voltage are easier to control in the experiment. Therefore, our predictions of dependence of the quantum transport quantified by the flux and tunneling as well as voltage can be tested in the experiment. 
By the interference techniques developed from quantum optics, researchers have began to have the control of the coherence. Our predictions of nontrivial dependence of the tunneling and coherence at fixed voltage, the flux and effective voltage at fixed coherence, as well as flux and coherence at fixed voltage, should be tested in the upcoming experiments.

Before leaving this subsection, the non-equilibriumness of quantum systems with non-resonance (shown in Fig.2(d)) will be discussed in detail in appendix B. As the flux decomposition has been carried out at the classical level, we will discuss the comparison of our quantum results to the classical limit in following section.

{\it The Eq.(\ref{20})-(\ref{23}) \& discussion above and the comparison to classical description will be carried out in next section and Eq.(\ref{25}). These, together with (\ref{29}) and (\ref{32}) on efficiency and non-equilibrium quantum thermodynamics, construct the main achievements (concepts) and theoretical framework put forward in this paper}.

\begin{figure}
\centering
$\begin{array}{cc}
 \includegraphics[width=1.6in,height=1.25in]{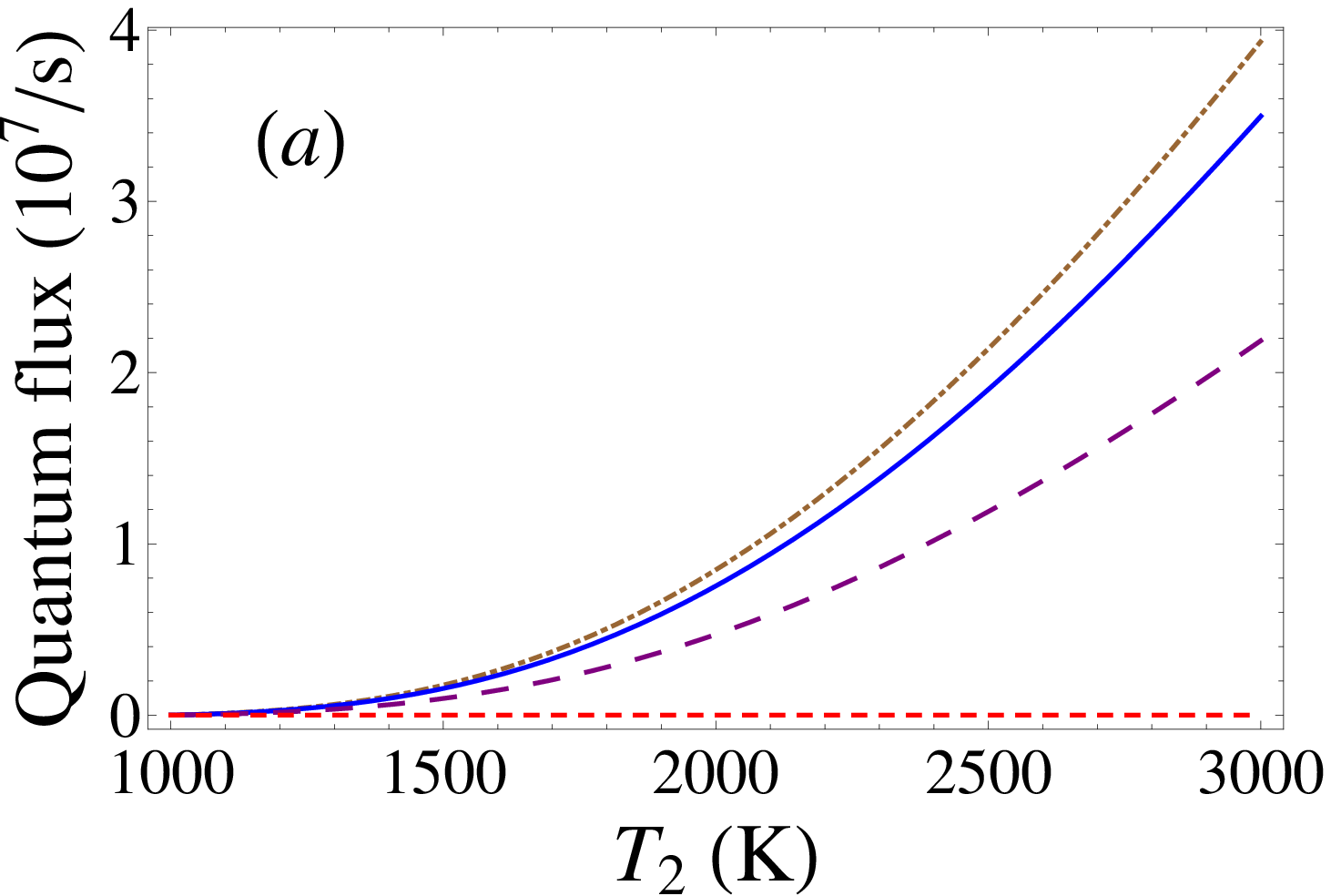}
&\includegraphics[width=1.6in,height=1.25in]{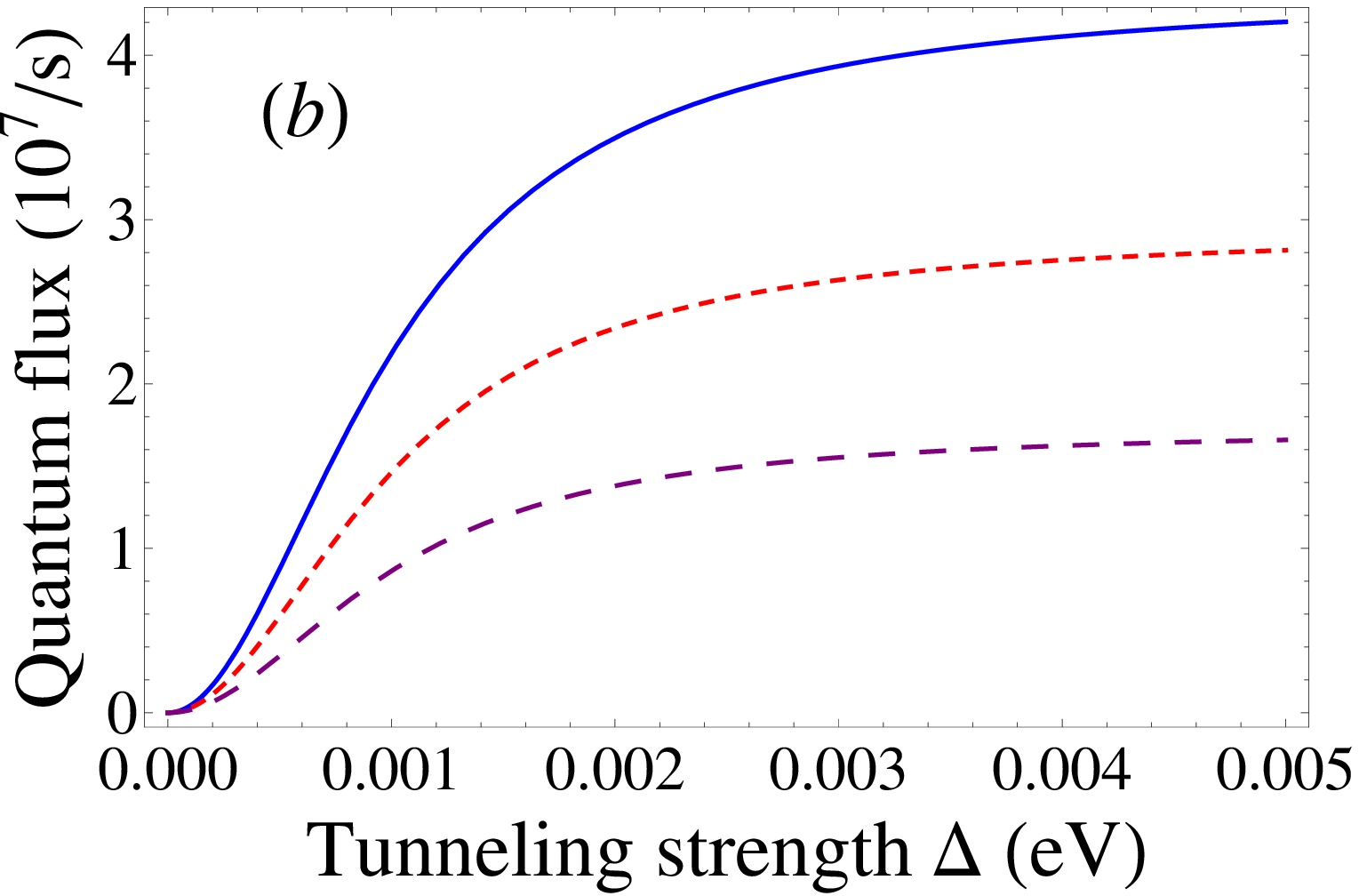}\\
\includegraphics[width=1.6in,height=1.25in]{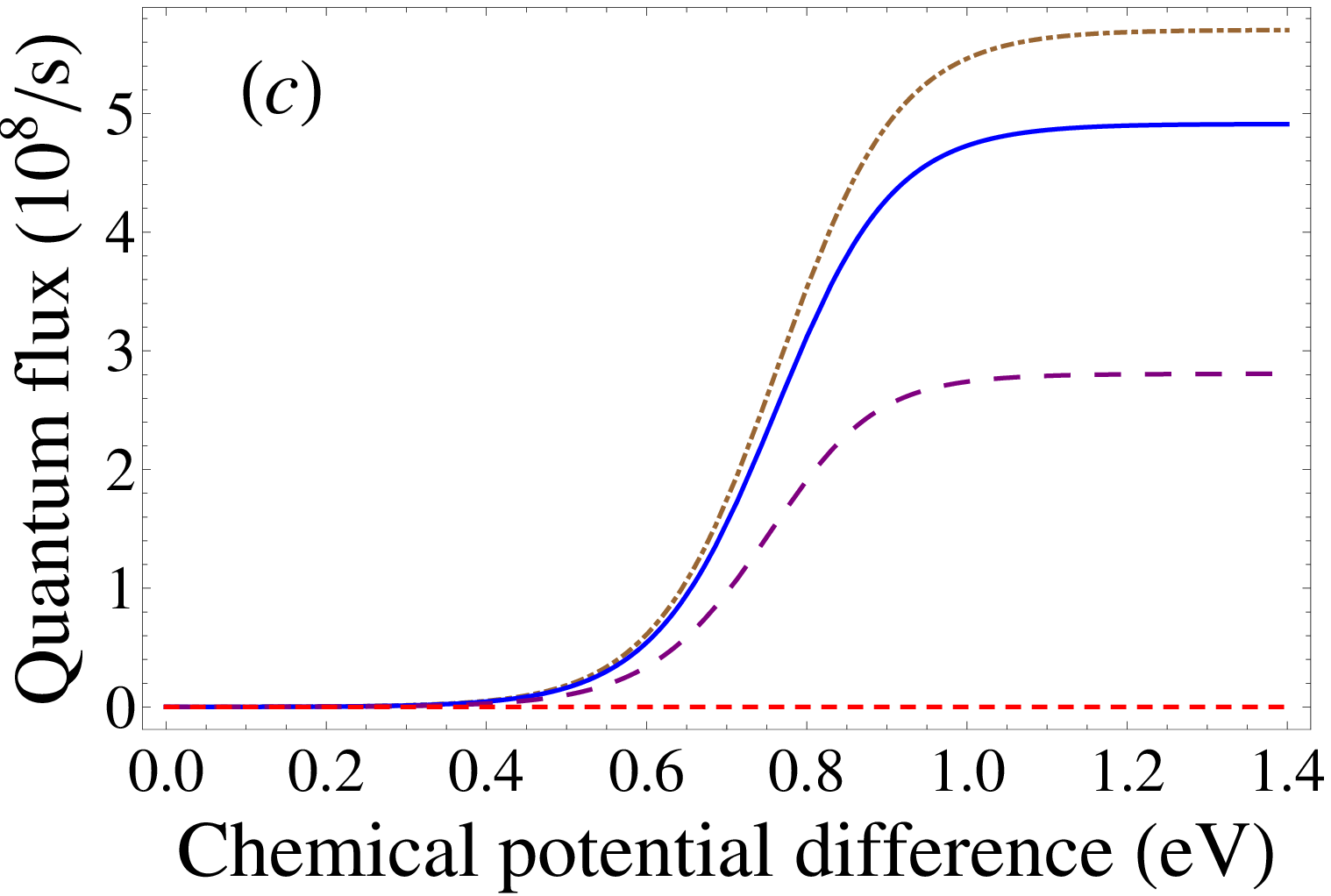}
&\includegraphics[width=1.6in,height=1.25in]{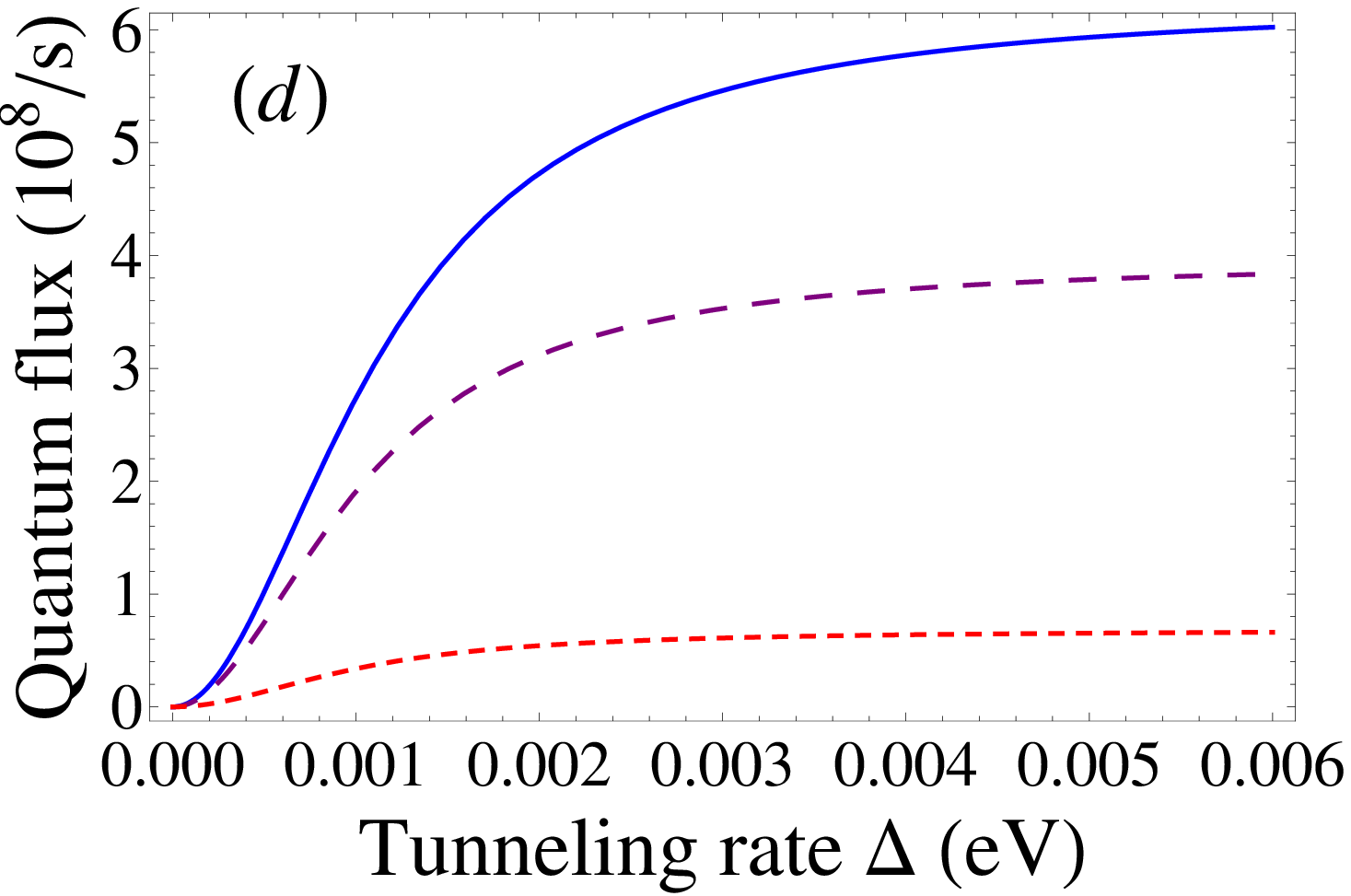}\\
\end{array}$
\caption{(Color online) Quantum flux varies for (a,b)bosonic and (c,d)fermionic reservoirs with (a,c)voltage and (b,d)tunneling strength. (a,c) Brown, blue, purple and red curves correspond to $\Delta=3$meV, $2$meV, $1$meV and 0, respectively; (b) Blue, red and purple curves correspond to $T_2=3000$K, 2650K and 2300K, respectively; (d) Blue, purple and red curves correspond to $\mu_2=1.0$eV, $0.8$eV and $0.6$eV, respectively. Standard parameters are $\varepsilon_1=0.798$eV, $\varepsilon_2=0.8$eV, $\lambda=21$cm$^{-1}$, (a,b) $T_1=1000$K and (c,d) $T=900$K, $\mu_1=0$}
\label{Fig.3}
\end{figure}

\section{Comparison with classical description}
Here we will discuss the classical correspondence. The non-trivial classical limit in this model is that (a) $\Omega\gg\omega$ and (b) {\it high temperature} where (a) keeps the non-vanishing transition rate and also effectively suppress the height of the barrier to ensure one energy level being closed to the top of barrier, (b) is somewhat equivalent to $\hbar\rightarrow 0$. Therefore the flux and coherence can be expanded in terms of $\omega/\Omega$ and $\varepsilon/k_B T$ ($\gamma\equiv\Gamma/\hbar^2$, $\Delta\equiv\hbar\Omega$)
\begin{equation}
\begin{split}
& \mathcal{J}_q^b=\frac{2\gamma}{3}\frac{T_2-T_1}{T_1+T_2}+{\cal O}\left(\frac{\varepsilon_1+\varepsilon_2}{k_B T_1},\frac{\varepsilon_1+\varepsilon_2}{k_B T_2},\frac{\omega}{\Omega}\right)\\[0.2cm]
& \mathcal{J}_q^f=\frac{\gamma}{6}\frac{\mu_2-\mu_1}{k_B T}+{\cal O}\left(\frac{\varepsilon_1-\mu_1}{k_B T},\frac{\varepsilon_2-\mu_2}{k_B T},\frac{\omega}{\Omega}\right)\\[0.2cm]
& |\textup{Im}\rho_{12}^b|=\frac{\gamma}{3\omega}\frac{T_2-T_1}{T_1+T_2}\left(\frac{\omega}{\Omega}\right)+{\cal O}\left(\frac{\varepsilon_1+\varepsilon_2}{k_B T_1},\frac{\varepsilon_1+\varepsilon_2}{k_B T_2},\frac{\omega^2}{\Omega^2}\right)\\[0.2cm]
& |\textup{Im}\rho_{12}^f|=\frac{\gamma}{12\omega}\frac{\mu_2-\mu_1}{k_B T}\left(\frac{\omega}{\Omega}\right)+{\cal O}\left(\frac{\varepsilon_1-\mu_1}{k_B T},\frac{\varepsilon_2-\mu_2}{k_B T},\frac{\omega^2}{\Omega^2}\right)
\end{split}
\label{24}
\end{equation}
The leading order term in flux is the classical correspondence where $\hbar$ disappeared and it is also proportional to the voltage. Quantum effect is attributed to the {\it higher order terms}. On the other hand, we can also see that {\it coherence effect comes in since the order of $\omega/\Omega$}


On the other hand, by the measurement of coherence the quantum flux shows a non-monotonic behavior as a function of coherence, according to Eq.(\ref{23}). 
This is due to the up-hill and down-hill behaviors of coherence as explained in detail from Fig.2(a) and 2(b) before. From this point, it should be noted that in open quantum systems the coherence does not always enhance the flux and  transport, but sometimes it can inhibit them, due to the mixture of classical behavior of motion.

Furthermore, as we know that classical flux monotonically increases as external voltage, shown in Eq.(\ref{24}). In quantum case, however, it can be clearly illustrated from Eq.(\ref{23}) that external voltage leads to the decrease of the flux for small tunneling and increase of the flux for large tunneling, by fixing the value of coherence as shown in Fig.2(c). The explanations of the behavior is given already in the previous subsection. In next section we will discuss the macroscopic quantum transport relevant to experiments.


\section{Quantum transport and non-equilibrium thermodynamics}

\subsection{Transfer Efficiency}

From the definition of our quantum flux we know that it provides a measurement on how much energy (chemical species) is transported from one site to another. Therefore the energy transfer efficiency ETE and chemical reaction efficiency (or charge transfer efficiency) CRE can be introduced in terms of flux, so that $\eta=\mathcal{J}_q/(\mathcal{J}_q+\mathcal{A}_{22}^{gg}\rho_{22})$. After some mathematical steps we have
\begin{equation}
\begin{split}
& \eta^b=\frac{\left(n_{\varepsilon}^{T_2}-n_{\varepsilon}^{T_1}\right)\frac{\Delta^2}{\hbar^2\omega^2}}{n_{\varepsilon}^{T_2}\left[B\left(T_1,T_2,\omega\right)+\left(\bar{n}_{\varepsilon}+2\right)\frac{\Delta^2}{\hbar^2\omega^2}\right]}\\[0.2cm]
& \eta^f=\frac{\left(n_{\varepsilon}^{\mu_2}-n_{\varepsilon}^{\mu_1}\right)\frac{\Delta^2}{\hbar^2\omega^2}}{n_{\varepsilon}^{\mu_2}\left[F\left(\mu_1,\mu_2,T,\omega\right)+\left(2-\bar{n}_{\varepsilon}\right)\frac{\Delta^2}{\hbar^2\omega^2}\right]}
\end{split}
\label{25}
\end{equation}
where the definition of two functions $B$ and $F$ are shown in Appendix C.
As shown in Fig.10 in Appendix, there are two plateaus in CRE, in contrast to flux. The second one is easy to understand which is due to the Pauli exclusion principle as the same as in flux, the reason for the first plateau is that at the beginning there is an improvement of CRE due to the non-vanishing flux in the non-equilibrium regime. As voltage from environments is below the excitation energy gap $\varepsilon_2$ the excitations absorbed by the molecular system is suppressed until reaching the gap, then it leads to an abrupt increase to another higher plateau. This is because of the significant improvement of excitations, based on Fermi-Dirac statistics.

In terms of the voltage (temperature difference for heat transport and chemical potential difference for chemical reactions) and quantum coherence, we can use Eq.(\ref{21}) and (\ref{25}) to eliminate the tunneling and then obtain the dependence of ETE and CRE (CTE) on voltage and coherence. In principle, the effect of coherence can be observed from the interference experiments, such as Hamburg-Brown-Twist setup \cite{Scully97}. Fig.5(a) and 5(b) collect the behavior of the ETE as well as the CRE as functions of coherence at several fixed voltages. For energy transport process in molecules, we found that by fixing the voltage, the increase of the coherence leads to a significant improvement of ETE while in the large tunneling regime ETE is significantly promoted by the reduction of the coherence. For the chemical reaction process and the charge transport in molecules, the coherence plays a crucial role on enhancing the CRE or CTE, as shown in Fig.5(b). In the large tunneling regime the influence of coherence is weak, since it approaches the quasi-classical regime. Furthermore, for both bosonic and fermionic reservoirs the voltage from external environments leads to further improvement of transfer efficiency, in addition to coherence.

In order to see how the quantum tunneling and environments affect the ETE and CRE through the bridge of coherence, the ETE and CRE as functions of $\Delta$ as well as voltage, for both bosonic and fermionic baths, are plotted in Fig.4. The tunneling strength and environments characterized by the effective non-equilibrium voltage
are shown to have competition on improving the ETE(CRE) and they can compensate for each other. Namely, the environments can lead to further enhancement when the hopping leads to saturation, and vice versa. Obviously, the optimization of ETE(CRE) cannot be achieved if any of those two aspects contributes too weakly.



\begin{figure}
\centering
$\begin{array}{cc}
 \includegraphics[width=1.6in,height=1.25in]{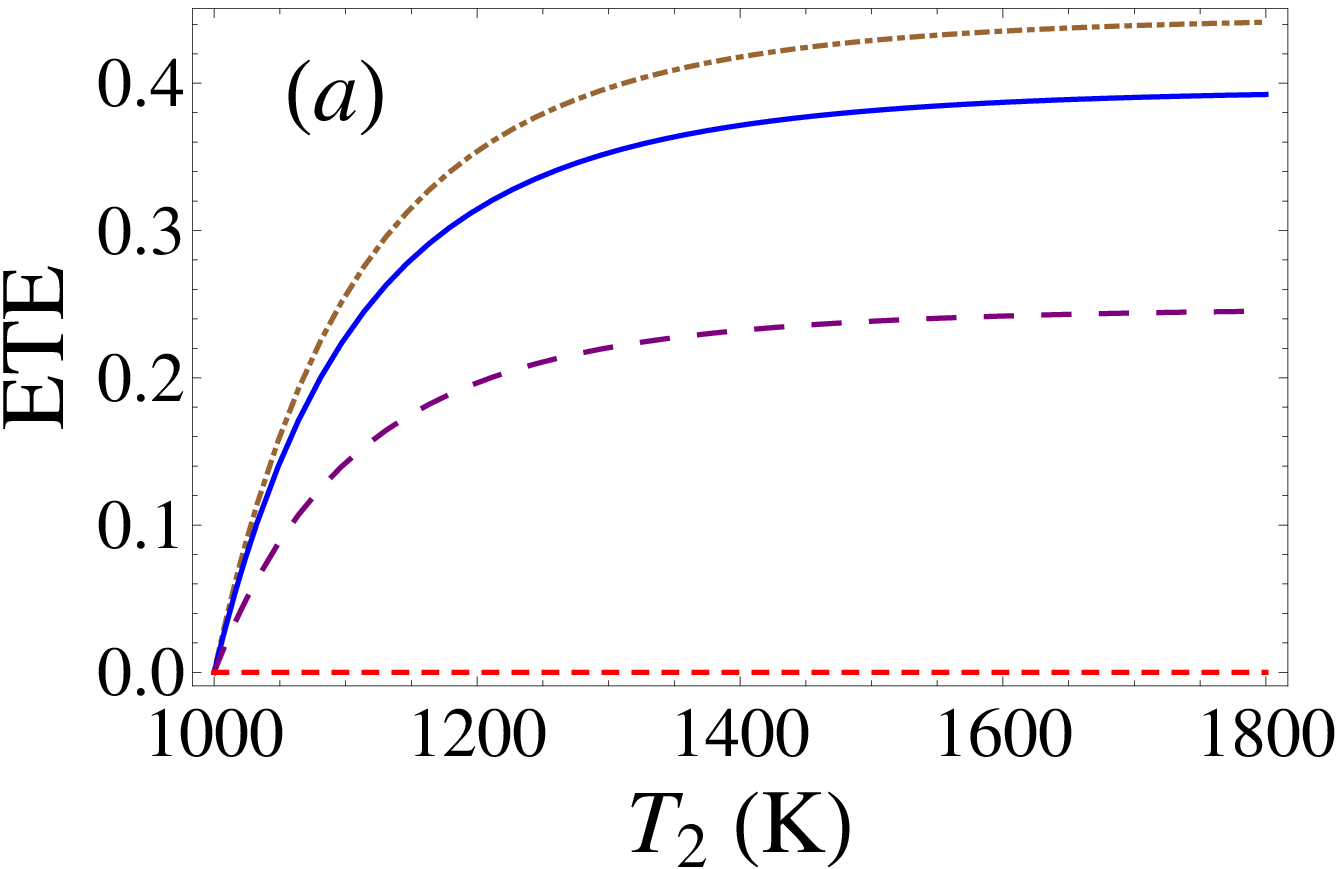}
&\includegraphics[width=1.6in,height=1.25in]{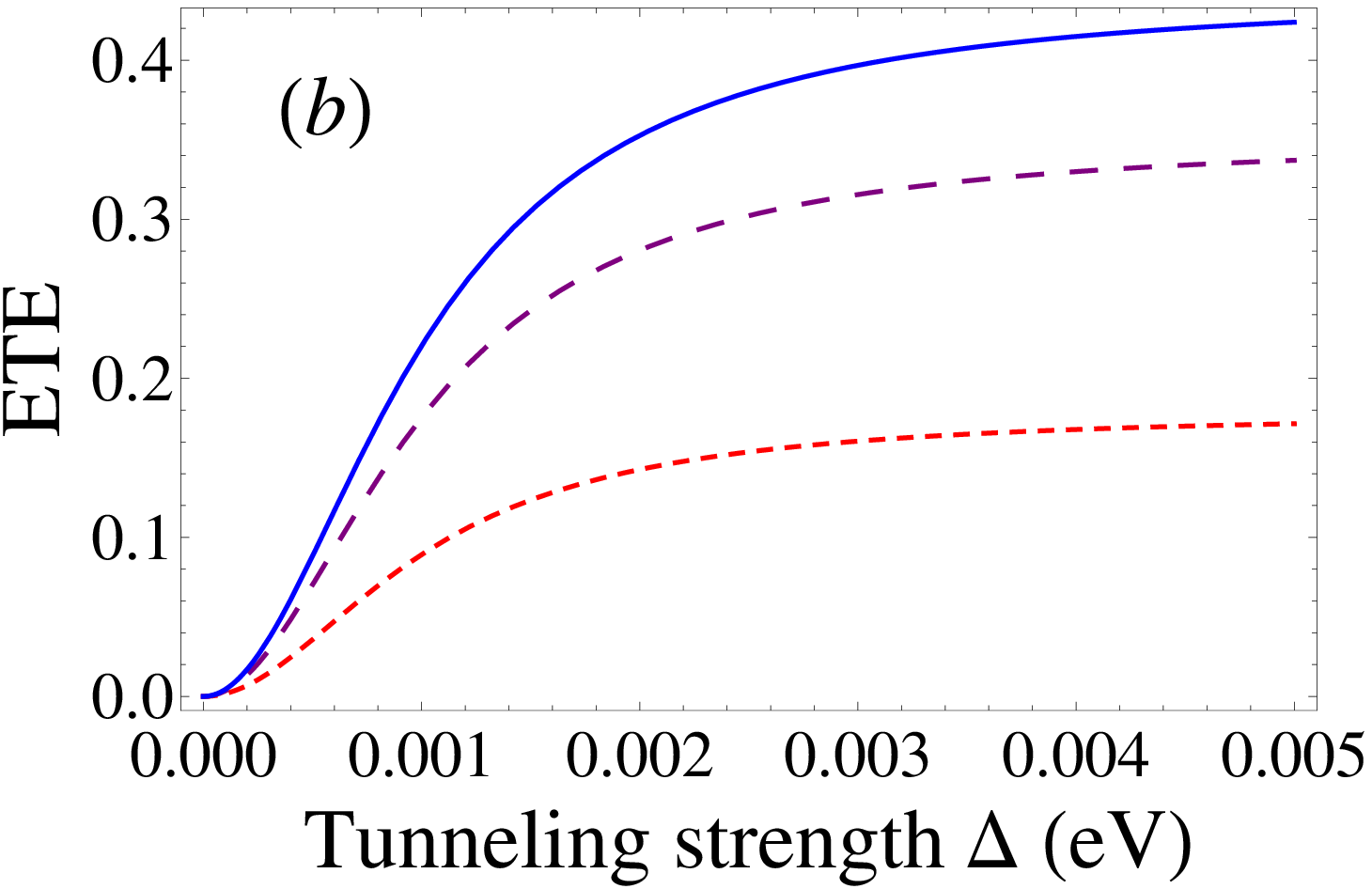}\\
\includegraphics[width=1.6in,height=1.25in]{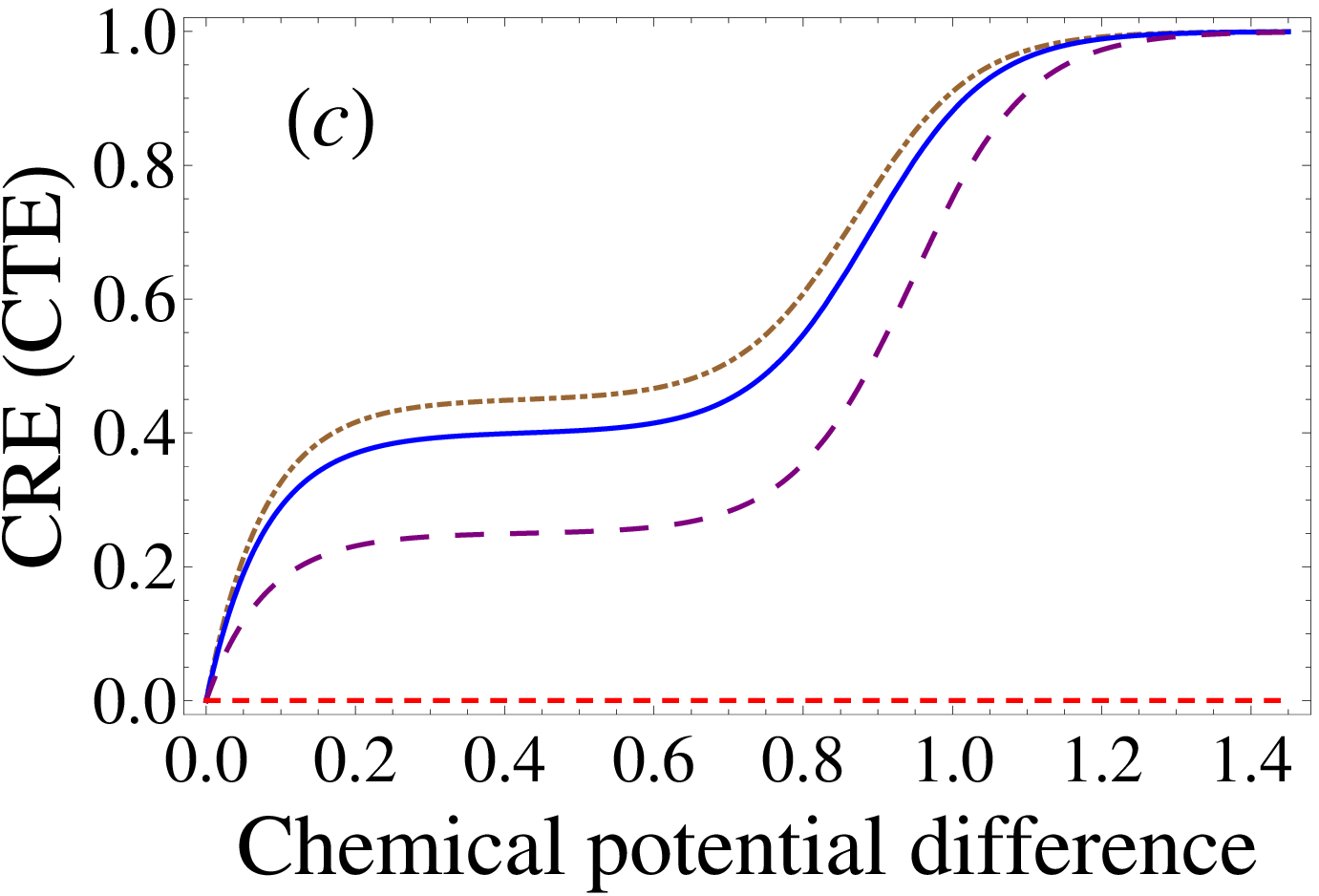}
&\includegraphics[width=1.6in,height=1.25in]{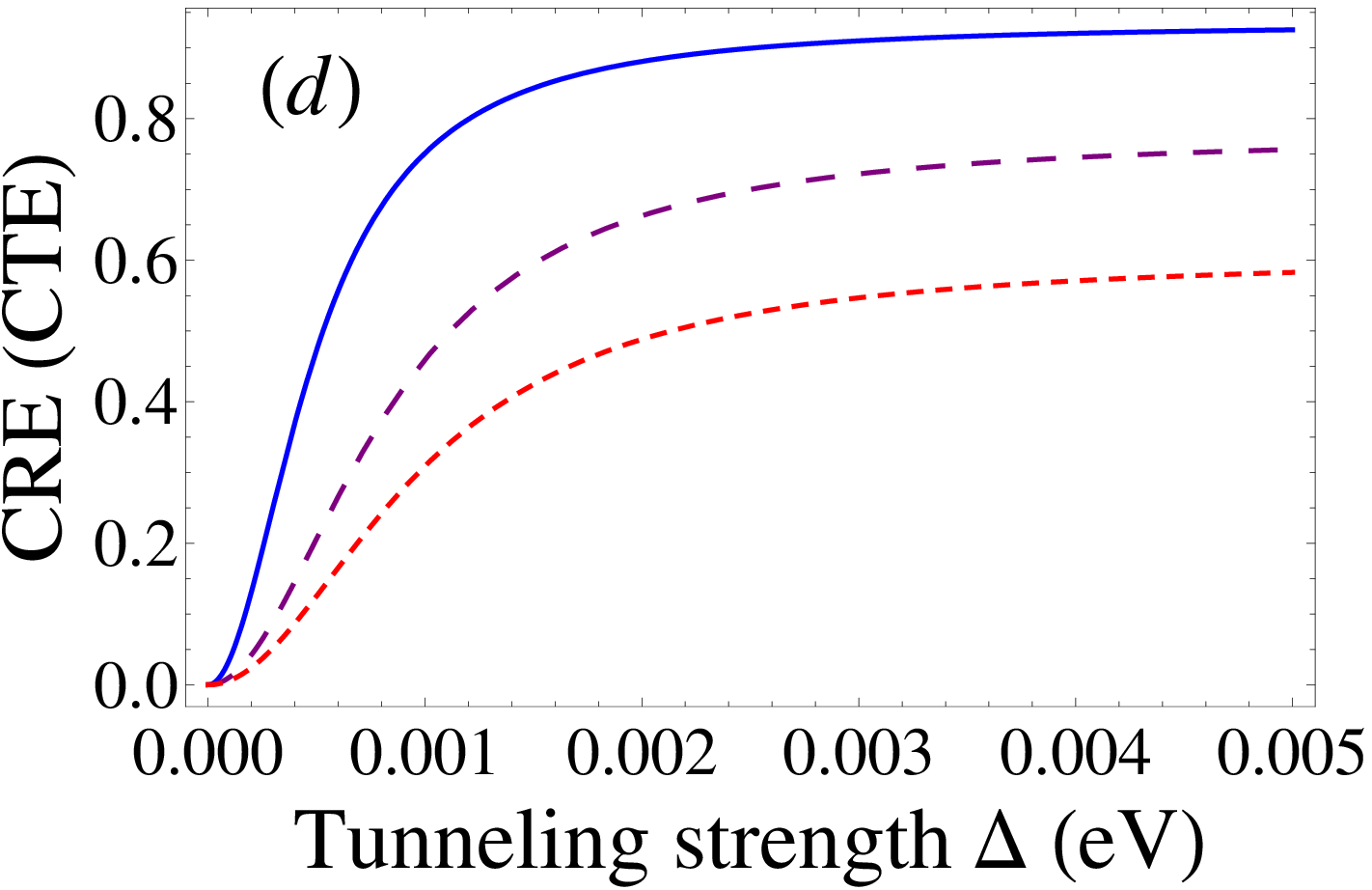}\\
\end{array}$
\caption{(Color online) (a,b)ETE and (c,d)CRE for bosonic and fermionic reservoirs, respectively, vary via tunneling strength and voltage. (a,c) Red, purple, blue and brown lines are for $\Delta=0$meV, 1meV, 2meV and 3meV, respectively; (b) Red, purple and blue lines are for $T_2=1050$K, 1150K and 1300K, respectively; (d) Blue, purple and red lines are for $\mu_2=1.00$eV, 0.87eV and 0.74eV, respectively. Standard parameters are $\varepsilon_1=0.798$eV, $\varepsilon_2=0.8$eV, $\lambda=21$cm$^{-1}$, (a,b) $T_1=1000$K and (c,d) $T=900$K, $\mu_1=0$}
\label{Fig.4}
\end{figure}

Moreover, we can also see that the transfer efficiency of quantum systems coupled to fermionic environments (CRE) is much better than that for being coupled to heat baths (ETE), since the optimization of CRE is almost a perfect value of 100$\%$ 
from Eq.(\ref{25}) by $\mu_2\rightarrow\infty$, while ETE's is only about $42\%$ in our model. On the other hand, this indicates that there is little dissipation-decay present in the transport (decay back to ground state in molecule coupled to the bath with higher chemical potential) when the open quantum system is at far-from-equilibrium, in the chemical reaction process. This kind of high efficiency of 70$\%$ was recently observed in the measurement of conductance of a ferrocene-based organometallic molecular wire \cite{Sita05}. In contrast, the heat dissipation is much larger in the quantum heat engine (QHE). This can be understood as follows: from the fermi distribution we know each mode of the reservoirs with higher chemical potential is fully occupied, thus Pauli exclusion principle causes the emission of one quasi-electron from the molecule back to high-chemical potential reservoir to be forbidden, hence almost all of the excitations are transported to other states. But for bosonic baths, the dissipation is unavoidable since emission of particles is always allowed, without the restriction by Pauli principle.

Finally, as we can see from the discussion above, the large voltage is often necessary and reasonable for optimizing the transport properties, such as ETE and CRE. In particular, in the light-harvesting complex, the radiation bath from the Sun which serves as an energy source is at 5870K \cite{JPCL2013}. The cooler surrounding, which is originated from the vibrations of proteins in chlorophyll, is at around 200$\sim$300K \cite{inter2014}. Consequently the temperature gradient becomes larger than 1000K.

\subsection{Macroscopic Currents and Energy Dissipation}

In the experiments, the observables provide direct measures of energy dissipation (heat current) and chemical current on the macroscopic level. Therefore we need to explore the connection of the quantum flux and voltage to these macroscopic quantities. First the total entropy production rate (EPR) is introduced $\dot{S}_t=k_B \mathcal{J}_q\textup{log}\frac{\mathcal{A}_{gg}^{22}\mathcal{A}_{11}^{gg}\mathcal{A}_{22}^{11}}{\mathcal{A}_{11}^{22}\mathcal{A}_{gg}^{11}\mathcal{A}_{gg}^{22}}$ where coherence effect has been already contained in matrix $\mathcal{A}$. Within the near-resonant approximation above, EPR reads

\begin{figure}
\centering
$\begin{array}{cc}
 \includegraphics[width=1.6in,height=1.2in]{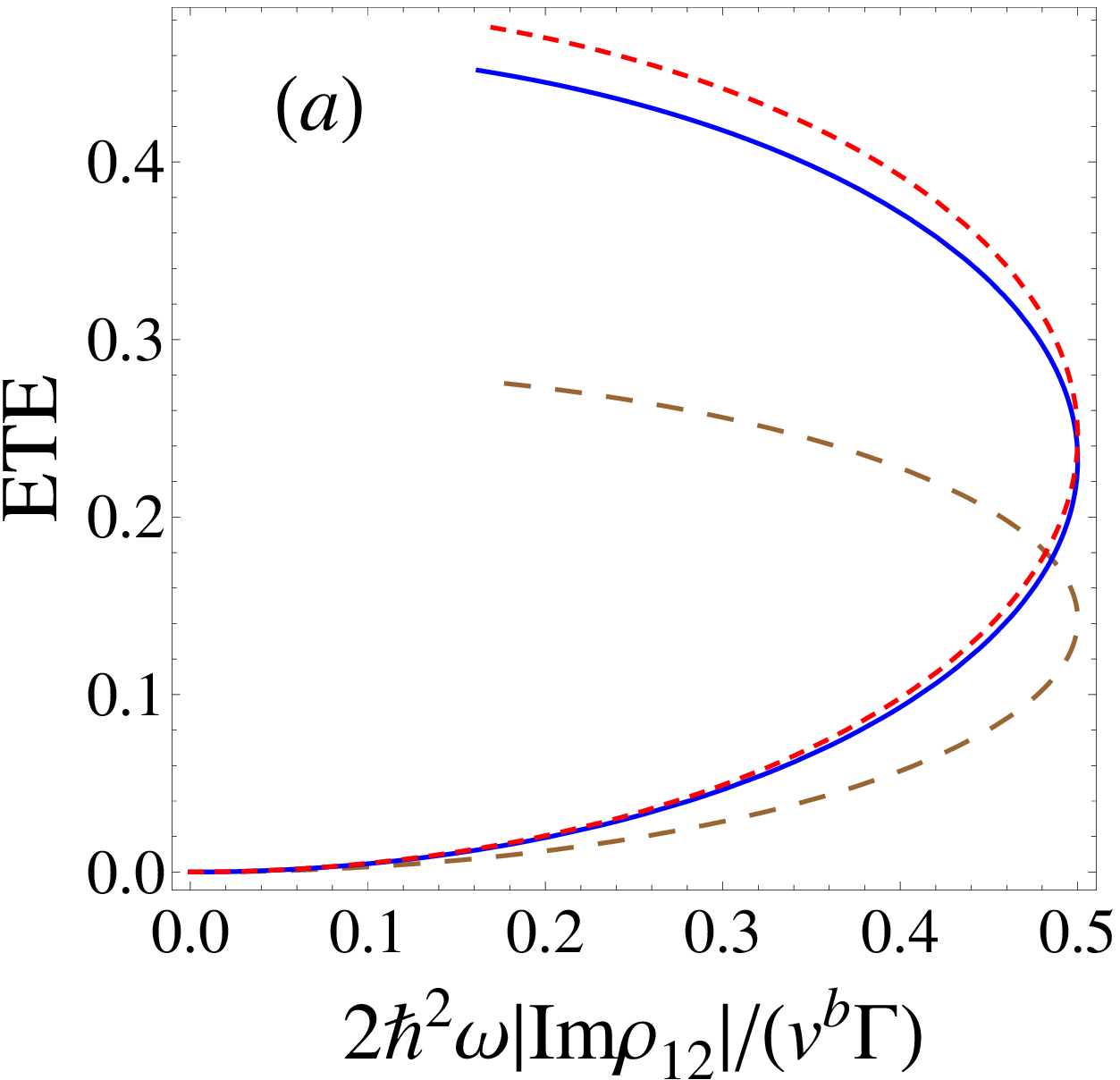}
&\includegraphics[width=1.6in,height=1.2in]{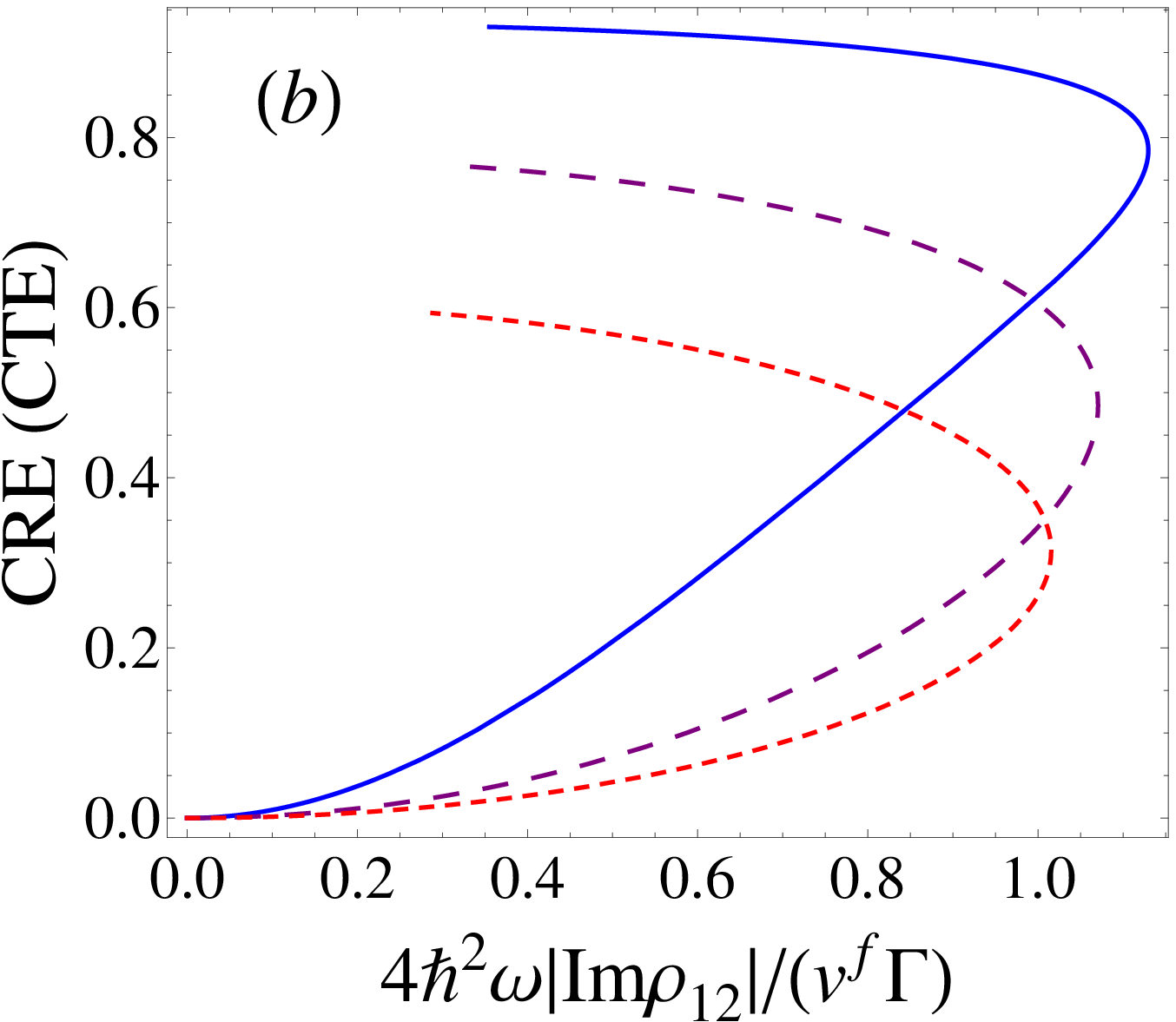}\\
\end{array}$
\caption{(Color online) ETE for bosons and CRE(CTE) for fermions vary via coherence, by fixing the voltage. (a) Brown, blue and red lines are for $T_2=1100$K, 1400K and 1800K, respectively; (b) Red, purple and blue lines are for $\mu_2=0.74$eV, 0.87eV and 1.0eV, respectively. Standard parameters are $\varepsilon_1=0.798$eV, $\varepsilon_2=0.8$eV, $\lambda=21$cm$^{-1}$, (a) $T_1=1000$K and (b) $T=900$K, $\mu_1=0$}
\label{Fig.5}
\end{figure}


\begin{equation}
\begin{split}
& \dot{S}_t^b=\frac{\varepsilon_1+\varepsilon_2}{2}\left(\frac{1}{T_1}-\frac{1}{T_2}\right)\mathcal{J}_q^b\\[0.2cm]
& \dot{S}_t^f=k_B\mathcal{J}_q^f\ \textup{log}\frac{n_{\varepsilon}^{\mu_2}(1-n_{\varepsilon}^{\mu_1})}{n_{\varepsilon}^{\mu_1}\left(1-n_{\varepsilon}^{\mu_2}-\frac{\Delta^2 a_1/\hbar^2\omega^2}{\sqrt{1+4\Delta^2/\hbar^2\omega^2}}\right)}
\end{split}
\label{27}
\end{equation}
where $a_1\equiv n_{\varepsilon_2}^{\mu_2}-n_{\varepsilon_1}^{\mu_1}-(n_{\varepsilon_1}^{\mu_2}-n_{\varepsilon_2}^{\mu_1})$ and $b$, $f$ correspond to bosonic and fermionic reservoirs, respectively. In heat transport, the $1^{\textup{st}}$ and $2^{\textup{nd}}$ laws in thermodynamics give
\begin{equation}
\dot{Q^b}_2-\dot{Q^b}_1-\dot{E}=0,\quad -\frac{\dot{Q^b}_2}{T_2}+\frac{\dot{Q^b}_1}{T_1}+\dot{S}=\dot{S}_t^b
\label{28}
\end{equation}
Here 1 or 2 refers to the site 1 or 2 with each coupled with different bosonic bath respectively. $Q$ denotes the magnitude of energy flowing into reservoir. Notice the entropy production rate $\dot{S}$ of system vanishes at steady state so that we have the energy dissipation by using Eq.(\ref{27}) and Eq.(\ref{28})
\begin{equation}
\dot{Q^b}_1=\frac{\varepsilon_1+\varepsilon_2}{2}\mathcal{J}_q^b
\label{29}
\end{equation}
which gives $\dot{Q^b}_1=-\frac{2\varepsilon t}{\hbar}\textup{Im}\rho_{12}$ in the limit $\theta\rightarrow -\frac{\pi}{2}$. This recovers the result in Ref.[39] so that our introduction of EPR for quantum steady state has its physical rational. In high temperature limit the asymptotic behaviors of EPR and energy dissipation are shown to be: $\dot{S}_t^b\sim (T_2-T_1)^2$ and $\dot{Q}_1\sim T_2-T_1$, which coincides with Fourier's law. For fermionic reservoirs, the chemical pumping is contributed by chemical flows, carried by the currents flowing into and out from the system

\begin{figure}
\centering
$\begin{array}{cc}
 \includegraphics[width=1.6in,height=1.23in]{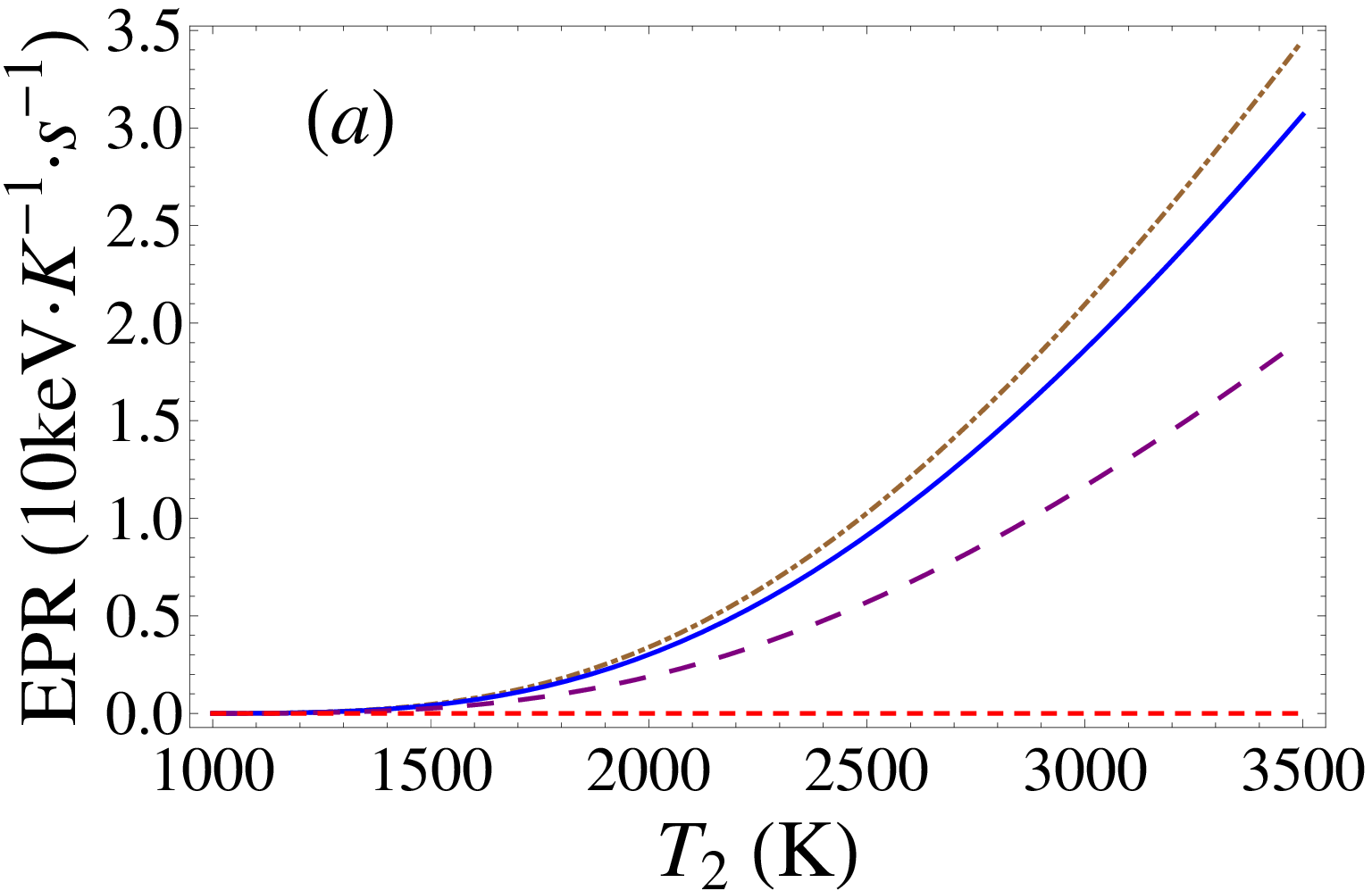}
&\includegraphics[width=1.6in,height=1.19in]{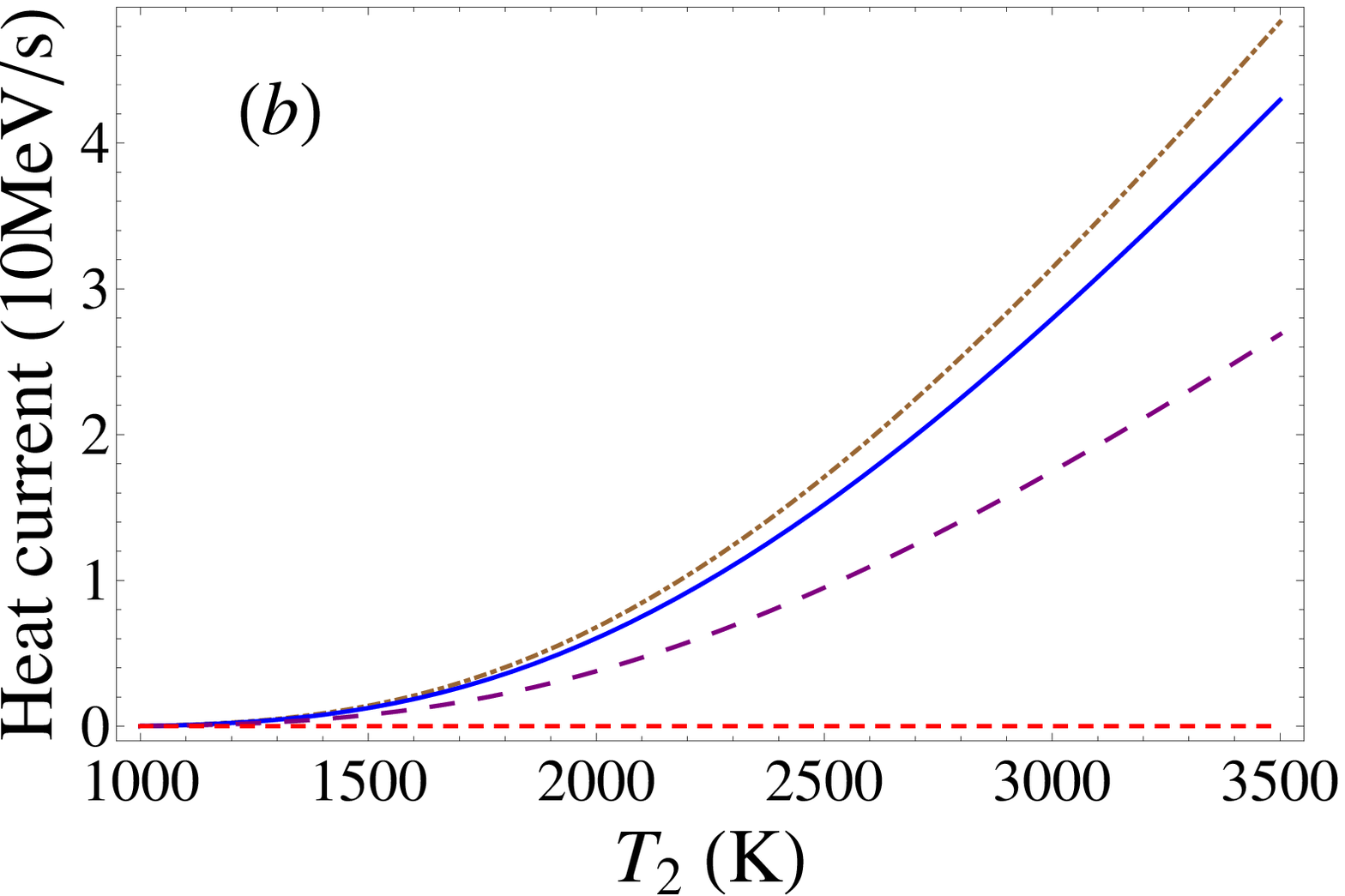}\\
\includegraphics[width=1.6in,height=1.3in]{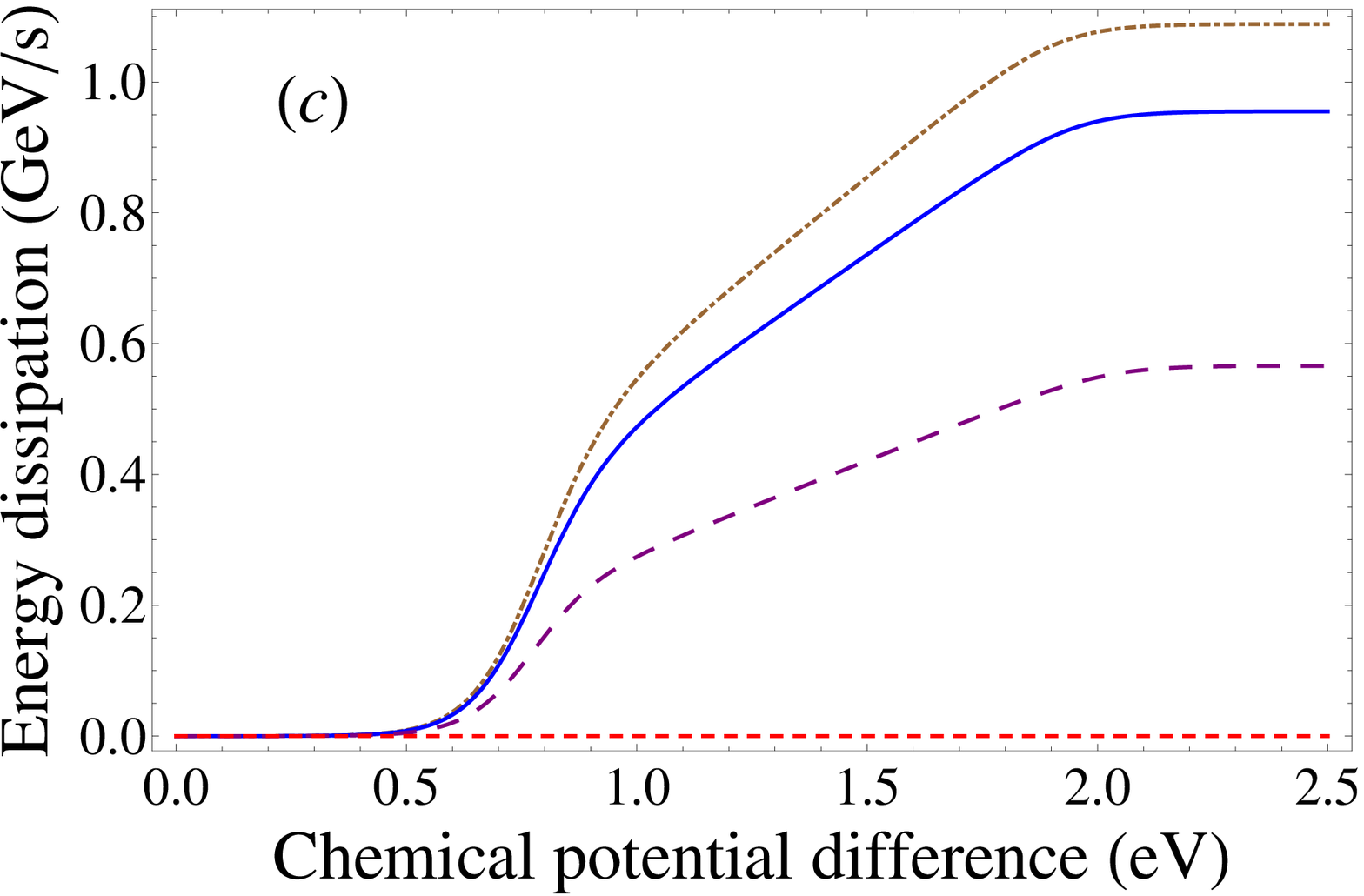}
&\includegraphics[width=1.6in,height=1.3in]{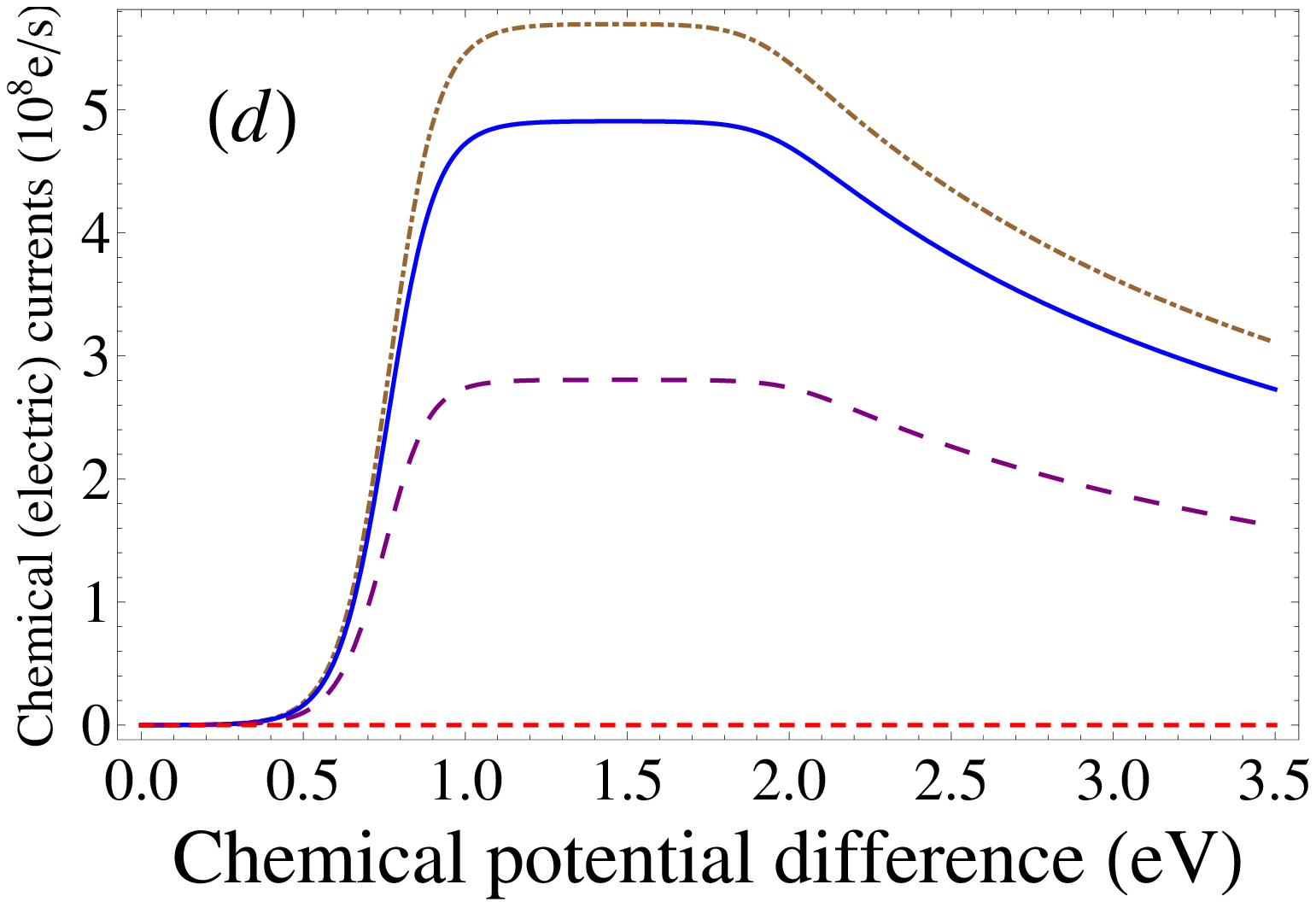}\\
\end{array}$
\caption{(Color online) (a) EPR and (b) Heat current vary as functions of temperature difference; (c) Energy dissipation and (d) Electric current vary as functions of chemical voltage. (a,b,c,d) Brown, blue, purple and red lines correspond to $\Delta=3$meV, 2meV, 1mev and 0, respectively. Standard parameters are $\varepsilon_1=0.798$eV, $\varepsilon_2=0.8$eV, $\lambda=21$cm$^{-1}$, $T_1=1000$K, $T=900$K and $\mu_1=0$}
\label{Fig.6}
\end{figure}


\begin{equation}
I_m^{(2)}-I_m^{(1)}=0,\quad \frac{\mu_2 I_m^{(2)}}{T}-\frac{\mu_1 I_m^{(1)}}{T}+\dot{S}=\dot{S}_t^f
\label{30}
\end{equation}
which leads to the chemical current at steady state
\begin{equation}
i_s=\frac{q k_B T}{\mu_2-\mu_1}\mathcal{J}_q^f\textup{log}\frac{n_{\varepsilon}^{\mu_2}(1-n_{\varepsilon}^{\mu_1})}{n_{\varepsilon}^{\mu_1}\left(1-n_{\varepsilon}^{\mu_2}-\frac{\Delta^2 a_1/\hbar^2\omega^2}{\sqrt{1+4\Delta^2/\hbar^2\omega^2}}\right)}
\label{31}
\end{equation}
where $i_s=q I_m^{(2)}$ and the energy dissipation reads
\begin{equation}
\begin{split}
\dot{Q}_1^f & =\mu_2I_m^{(2)}-\mu_1I_m^{(1)}\\[0.2cm]
& =k_B T\mathcal{J}_q^f\textup{log}\frac{n_{\varepsilon}^{\mu_2}(1-n_{\varepsilon}^{\mu_1})}{n_{\varepsilon}^{\mu_1}\left(1-n_{\varepsilon}^{\mu_2}-\frac{\Delta^2 a_1/\hbar^2\omega^2}{\sqrt{1+4\Delta^2/\hbar^2\omega^2}}\right)}
\end{split}
\label{32}
\end{equation}
Notice that the mathematical forms and analysis for system coupled to fermionic reservoirs Eq.(\ref{20}), (\ref{25}), (\ref{30})-(\ref{32}) can also be directly applied to charge transport in single molecules, i.e. electric current with $I-V$ relationship, where $i_e=eI_m$, which will be addressed later on the correlation to the experiments. Eq.(\ref{27}), (\ref{29}), (\ref{31}) and (\ref{32}) show that the non-equilibrium quantum flux serves as a driving force for the macroscopic energy dissipation and chemical (electric) current directly measured in experiments.The physical currents are generated and detailed balance condition is broken when the energy pump emerges ($T_1\neq T_2$ or $\mu_1\neq \mu_2$).
This furthermore reveals the robustness of the connection between non-equilibriumness and quantum transport and provides a measurement on how non-equilibriumness controls the transport properties.

Fig.6 collects the voltage dependence of transport coupled to both bosonic and fermionic environments. The heat current shows a monotonic increase with respect to temperature difference, which is reasonable due to the large energy pumping with increasing temperature and large dissipation at far-from-equilibrium. Due to the Pauli principle for fermions in recombination dissociation reactions in chemical process, there is an upper limit for the pumping work at steady state illustrated in Fig.6(c). 
Thus the chemical current will drop at high voltage (shown by large $\mu_2$ limit in Eq.(\ref{31}) and Fig.6(d)). This has been observed for electric current in the experiments on $I-V$ curve of electron transfer in single molecules \cite{Tao08}.

\begin{figure}
\centering
$\begin{array}{cc}
 \includegraphics[width=1.6in,height=1.19in]{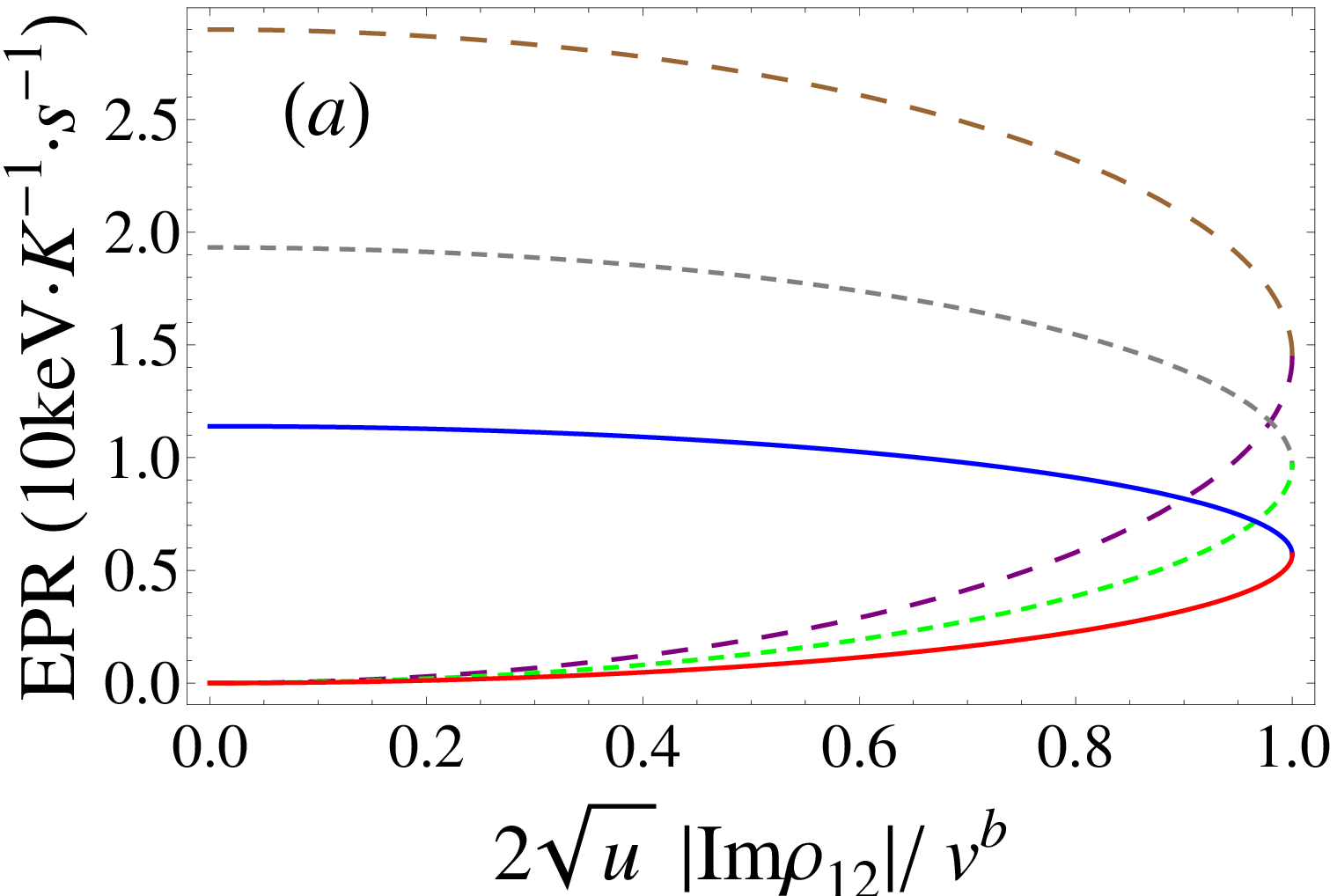}
&\includegraphics[width=1.65in,height=1.3in]{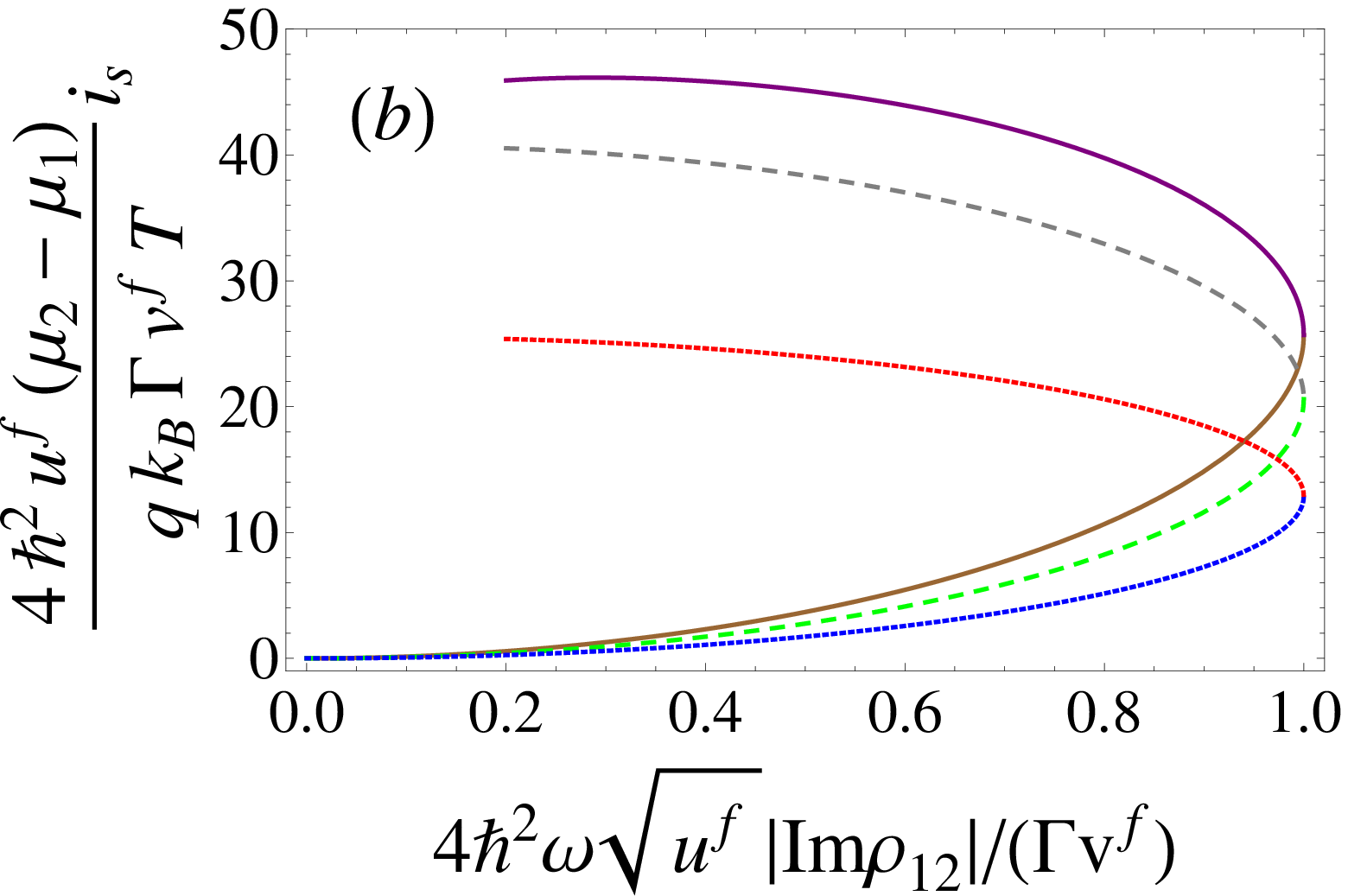}\\
\end{array}$
\caption{(Color online) (a) EPR for energy transport and (b) chemical (electric) current for chemical reaction (charge transport) vary as functions of coherence; (a) Large dashed, medium dashed and solid lines are for $T_2=3200$K, 2850K and 2500K, respectively; (b) Tiny dashed, medium dashed and solid lines are for $\mu_2=1.0$eV, 1.6eV and 2.2eV, respectively. Standard parameters are $\varepsilon_1=0.798$eV, $\varepsilon_2=0.8$eV, $\lambda=21$cm$^{-1}$, $T_1=1000$K, $T=900$K and $\mu_1=0$}
\label{Fig.7}
\end{figure}

To see the coherence effect on the macroscopic EPR, heat current, chemical (electric) current and energy dissipation, we apply Eq.(\ref{20})-(\ref{23}) to Eq.(\ref{27}), (\ref{29}), (\ref{31}) and (\ref{32}) by eliminating the tunneling $\Delta$, similar to the transfer efficiency studied in the previous subsection. These behaviors are shown in Fig.7(a) and 7(b) by fixing voltages, for energy transport and chemical reaction (charge transport), respectively. Due to the fixed voltage, those macroscopic observables are different from each other up to just a scaled factor, so that we only display the EPR and chemical (electric) current here. As shown in Fig.7(a) and 7(b), the non-monotonic behaviors of those macroscopic observables in terms of the coherence indicate that the coherence does not always promote the transport. In the large tunneling regime coherence inhibits the quantum transport. This is distinct from the behavior of these macroscopic observables with respect to tunneling which always enhances the quantum transport. The non-monotonic behaviors in Fig.7 are due to the non-monotonic dependence of the coherence with respect to the tunneling, as discussed in the flux section.

Before leaving this section, we also calculate the dynamical decay rate of coherence, for both energy transport and chemical reaction processes. Both of them shows that the decay rates increase as the systems become far from equilibrium. However the behaviors with respect to tunneling uncover some novelty which is shown in detail in next section.

\section{Dynamical decay rate}
In this section we calculate the decay rate of the dynamical coherence (in time) in addition to steady state coherence, by the diagonalization  of matrix $\mathcal{M}$. It is obvious that the eigenvalues $\nu$ of $\mathcal{M}$ is complex and the real part governs the decay. Here we study the eigenvalue with largest modulus.

In general, we observed that the decay of the dynamical coherence in time is faster when the effective voltage or chemical potential measuring the non-equilibriumness away from the equilibrium increases. This implies that the large non-equilibrium driving forces can help to destroy the dynamical quantum coherence as shown in Fig.8(a) and 8(c). On the other hand, as the tunneling increases, the decay of the dynamical quantum coherence also increases. This implies
that the large quantum tunneling can help to destroy the dynamical coherence as shown in Fig.8(b) and 8(d), as the large tunneling reaches the classical limit.
\begin{figure}
\centering\includegraphics[width=3.46in,height=2.8in]{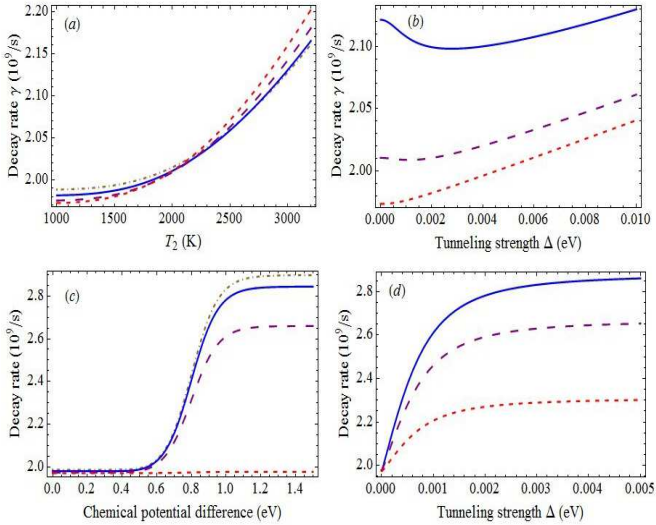}
\caption{(Color online) Dynamical decay rate varies as functions of voltage and tunneling strength, for (a,b)energy transfer and (c,d)chemical reaction processes. (a,c) Brown, blue, purple and red curves are for $\Delta=3$meV, 2meV, 1meV and 0, respectively; (b) Blue, purple and red lines are for $T_2=2800$K, 2000K and 1200K, respectively; (d) Blue, purple and red lines are for $\mu_2=1.0$eV, 0.87eV and 0.75eV, respectively; Standard parameters are $\varepsilon_1=0.798$eV, $\varepsilon_2=0.8$eV, $\lambda=21$cm$^{-1}$, $T_1=1000$K, $T=900$K and $\mu_1=0$}
\label{Fig.8}
\end{figure}

Fig.8(b) shows that a minimum of decay rate occurs when raising the temperature difference in the energy transport in single molecules. This is because of the improvement of interference effect by increasing voltage, which causes the decay of dynamical coherence to slow down at beginning of increasing tunneling. The decay is strengthened for large tunneling owing to the quasi-classical limit.

\section{Conclusion}
In this work, we systematically developed the concept and quantification of curl flux for non-equilibrium quantum processes at steady state. The curl quantum flux measures the degree of non-equilibriumness via detailed balance breaking and time-irreversibility. It also reflects the degree of quantum coherence. We further applied our theoretical framework to the quantum transport in energy (charge) transfer in single molecules and the chemical recombination dissociation reactions. 
More significantly, the quantum flux is also sensitively affected by coherence which could be observed by quantum interference experiments. The coherence leads to the non-monotonic behavior of the flux, depending on the magnitude of quantum mechanical tunneling. 
Furthermore we investigated quantum transport and thermodynamics of the system in terms of our quantum flux. We found that the non-equilibrium quantum flux serves as an intrinsic driving force for the macroscopic observables such as currents in quantum transport. 
{\it These are the main innovation and achievements in
this paper, mathematically illustrated in Eq.(\ref{20})-(\ref{23}), Eq. (\ref{25}), Eq. (\ref{27}), Eq.(\ref{29}), Eq.(\ref{31}) and Eq. (\ref{32})}.

On the other hand, we also investigated the environmental effects governed by voltage, on quantum flux, namely, the non-equilibriumness and quantum transport. For energy transport in molecules, we show that the significant enhancement of quantum coherence and flux as well as quantum transport is achieved at large temperature difference of the two environmental baths. This indicated the far-from-equilibrium instead of conventional near-to-equilibrium condition is essential for the perfect energy transport and high quality of quantum heat engines. For chemical reactions and charge transport, in contrast, there is a saturation plateau for coherence and flux as well as quantum transport, although significant enhancement of them was already demonstrated earlier in this work. Therefore the peak and decrease of chemical (electric) current from our theoretical calculations was already observed in recent experiments on electron transfer in single molecules \cite{Tao08}.

Our theoretical framework in the article can be applied and generalized to the energy transfer in photosynthetic reaction and the description of electron transport in molecules and chemical reactions, which we will pursue in the future work.

\begin{acknowledgements}
Z. D. Zhang and J. Wang would like to thank the grant NSF-MCB-0947767 for supports.
\end{acknowledgements}

\appendix
\section{}
The super operator in Eq.(6) in main text take the form of
\begin{widetext}
\begin{equation}
\begin{aligned}
\mathcal{D}\left(\rho_S\right)= & -\frac{1}{2\hbar^2}\Bigg\{\left[\left(n_{\omega'_{1g}}^{T_1}+1\right)\Gamma_1\textup{cos}^2\frac{\theta}{2}+\left(n_{\omega'_{2g}}^{T_1}+1\right)\Gamma_2\textup{sin}^2\frac{\theta}{2}\right]
\sigma_{1g}^+\sigma_{1g}^-\rho_S+\frac{\textup{sin}\theta}{2}\left[\left(n_{\omega'_{1g}}^{T_2}+1\right)\Gamma_1-\left(n_{\omega'_{2g}}^{T_2}+1\right)\Gamma_2\right]\sigma_{1g}^+\sigma_{2g}^-\rho_S\\[0.24cm]
& +\frac{\textup{sin}\theta}{2}\left[\left(n_{\omega'_{1g}}^{T_1}+1\right)\Gamma_1-\left(n_{\omega'_{2g}}^{T_1}+1\right)\Gamma_2\right]\sigma_{2g}^+\sigma_{1g}^-\rho_S+\left[\left(n_{\omega'_{1g}}^{T_2}+1\right)\Gamma_1\textup{sin}^2\frac{\theta}{2}+\left(n_{\omega'_{2g}}^{T_2}+1\right)\Gamma_2\textup{cos}^2\frac{\theta}{2}\right]\sigma_{2g}^+\sigma_{2g}^-\rho_S\\[0.24cm]
& -\left[\left(n_{\omega'_{1g}}^{T_1}+1\right)\Gamma_1\textup{cos}^2\frac{\theta}{2}+\left(n_{\omega'_{2g}}^{T_1}+1\right)\Gamma_2\textup{sin}^2\frac{\theta}{2}\right]\sigma_{1g}^-\rho_S\sigma_{1g}^+-\frac{\textup{sin}\theta}{2}\left[\left(\bar{n}_{\omega'_{1g}}+2\right)\Gamma_1-\left(\bar{n}_{\omega'_{2g}}+2\right)\Gamma_2\right]\sigma_{1g}^-\rho_S\sigma_{2g}^+\\[0.24cm]
& -\left[\left(n_{\omega'_{1g}}^{T_2}+1\right)\Gamma_1\textup{sin}^2\frac{\theta}{2}+\left(n_{\omega'_{2g}}^{T_2}+1\right)\Gamma_2\textup{cos}^2\frac{\theta}{2}\right]\sigma_{2g}^-\rho_S\sigma_{2g}^++h.c.\\[0.24cm]
& +\left[\left(n_{\omega'_{1g}}^{T_1}\textup{cos}^2\frac{\theta}{2}+n_{\omega'_{1g}}^{T_2}\textup{sin}^2\frac{\theta}{2}\right)\Gamma_1\textup{cos}^2\frac{\theta}{2}+\left(n_{\omega'_{2g}}^{T_1}\textup{sin}^2\frac{\theta}{2}+n_{\omega'_{2g}}^{T_2}\textup{cos}^2\frac{\theta}{2}\right)\Gamma_2\textup{sin}^2\frac{\theta}{2}\right]\sigma_{1g}^-\sigma_{1g}^+\rho_S\\[0.24cm]
& +\frac{\textup{sin}\theta}{2}\left[\left(n_{\omega'_{1g}}^{T_1}\textup{cos}^2\frac{\theta}{2}+n_{\omega'_{1g}}^{T_2}\textup{sin}^2\frac{\theta}{2}\right)\Gamma_1-\left(n_{\omega'_{2g}}^{T_1}\textup{sin}^2\frac{\theta}{2}+n_{\omega'_{2g}}^{T_2}\textup{cos}^2\frac{\theta}{2}\right)\Gamma_2\right]\left(\sigma_{1g}^-\sigma_{2g}^+\rho_S+\sigma_{2g}^-\sigma_{1g}^+\rho_S\right)\\[0.24cm]
& +\left[\left(n_{\omega'_{1g}}^{T_1}\textup{cos}^2\frac{\theta}{2}+n_{\omega'_{1g}}^{T_2}\textup{sin}^2\frac{\theta}{2}\right)\Gamma_1\textup{sin}^2\frac{\theta}{2}+\left(n_{\omega'_{2g}}^{T_1}\textup{sin}^2\frac{\theta}{2}+n_{\omega'_{2g}}^{T_2}\textup{cos}^2\frac{\theta}{2}\right)\Gamma_2\textup{cos}^2\frac{\theta}{2}\right]\sigma_{2g}^-\sigma_{2g}^+\rho_S\\[0.24cm]
& -\left(n_{\omega'_{1g}}^{T_1}\Gamma_1\textup{cos}^2\frac{\theta}{2}+n_{\omega'_{2g}}^{T_1}\Gamma_2\textup{sin}^2\frac{\theta}{2}\right)\sigma_{1g}^+\rho_S\sigma_{1g}^--\frac{\textup{sin}\theta}{2}\left[\left(n_{\omega'_{1g}}^{T_1}+n_{\omega'_{1g}}^{T_2}\right)\Gamma_1-\left(n_{\omega'_{2g}}^{T_1}+n_{\omega'_{2g}}^{T_2}\right)\Gamma_2\right]\sigma_{1g}^+\rho_S\sigma_{2g}^-\\[0.24cm]
& -\left(n_{\omega'_{1g}}^{T_2}\Gamma_1\textup{sin}^2\frac{\theta}{2}+n_{\omega'_{2g}}^{T_2}\Gamma_2\textup{cos}^2\frac{\theta}{2}\right)\sigma_{2g}^+\rho_S\sigma_{2g}^-+h.c.\Bigg\}
\end{aligned}
\end{equation}
\end{widetext}
Under the resonance approximation, namely, $\Delta\ll\textup{min}(\varepsilon_1,\varepsilon_2)$, the occupation will be replaced by average value: $n_{\omega'_{ag}}^{T}\simeq n_{\varepsilon}^{T}\equiv\frac{1}{2}(n_{\omega'_{1g}}^T+n_{\omega'_{2g}}^T)$ for bosons and $n_{\omega'_{ag}}^{\mu}\simeq n_{\varepsilon}^{\mu}\equiv\frac{1}{2}(n_{\omega'_{1g}}^{\mu}+n_{\omega'_{2g}}^{\mu})$ for fermions. $\Gamma_1\simeq\Gamma_2\simeq\Gamma\equiv\frac{1}{2}\left(\Gamma_1+\Gamma_2\right)$. Hence the expressions for matrix elements of $\mathcal{M}$ can be written out
\begin{widetext}
\begin{equation}
\begin{aligned}
& \mathcal{M}_{gg}^{gg}=-\frac{2\Gamma}{\hbar^2}\left(n_{\varepsilon}^{T_1}+n_{\varepsilon}^{T_2}\right),\quad \mathcal{M}_{11}^{gg}=\frac{2\Gamma}{\hbar^2}\left(n_{\varepsilon}^{T_1}+1\right),\quad \mathcal{M}_{22}^{gg}=\frac{2\Gamma}{\hbar^2}\left(n_{\varepsilon}^{T_2}+1\right),\quad \mathcal{M}_{12}^{gg}=\mathcal{M}_{21}^{gg}=0\\
& \mathcal{M}_{gg}^{11}=\frac{2\Gamma}{\hbar^2}n_{\varepsilon}^{T_1},\quad \mathcal{M}_{11}^{11}=-\frac{2\Gamma}{\hbar^2}\left(n_{\varepsilon}^{T_1}+1\right),\quad \mathcal{M}_{22}^{11}=\mathcal{M}_{11}^{22}=0,\quad \mathcal{M}_{12}^{11}=-\frac{i\omega}{2}\textup{tan}\theta\\
& \mathcal{M}_{21}^{11}=\frac{i\omega}{2}\textup{tan}\theta,\quad \mathcal{M}_{gg}^{22}=\frac{2\Gamma}{\hbar^2}n_{\varepsilon}^{T_2},\quad \mathcal{M}_{22}^{22}=-\frac{2\Gamma}{\hbar^2}\left(n_{\varepsilon}^{T_2}+1\right),\quad \mathcal{M}_{12}^{22}=\frac{i\omega}{2}\textup{tan}\theta\\
& \mathcal{M}_{21}^{22}=-\frac{i\omega}{2}\textup{tan}\theta,\quad \mathcal{M}_{gg}^{12}=0,\quad \mathcal{M}_{11}^{12}=-\frac{i\omega}{2}\textup{tan}\theta,\quad \mathcal{M}_{22}^{12}=\frac{i\omega}{2}\textup{tan}\theta\\
& \mathcal{M}_{12}^{12}=i\omega-\frac{\Gamma}{\hbar^2}\left(n_{\varepsilon}^{T_1}+n_{\varepsilon}^{T_2}+2\right),\quad \mathcal{M}_{21}^{12}=0
\end{aligned}
\label{B.2}
\end{equation}
\end{widetext}
which are for energy transport in molecules, coupled to bonsonic baths. Then we have the expressions for matrix elements of $\mathcal{A}$
\begin{widetext}
\begin{equation}
\begin{aligned}
& \mathcal{A}_{gg}^{gg}=-\frac{2\Gamma}{\hbar^2}\left(n_{\varepsilon}^{T_1}+n_{\varepsilon}^{T_2}\right),\quad \mathcal{A}_{11}^{gg}=\frac{2\Gamma}{\hbar^2}\left(n_{\varepsilon}^{T_1}+1\right),\quad \mathcal{A}_{22}^{gg}=\frac{2\Gamma}{\hbar^2}\left(n_{\varepsilon}^{T_2}+1\right),\quad \mathcal{A}_{gg}^{11}=\frac{2\Gamma}{\hbar^2}n_{\varepsilon}^{T_1}\\[0.22cm]
& \mathcal{A}_{11}^{11}=-\frac{2\Gamma}{\hbar^2}\left(n_{\varepsilon}^{T_1}+1\right)-\frac{\frac{\omega^2\Gamma}{2\hbar^2}\left(n_{\varepsilon}^{T_1}+n_{\varepsilon}^{T_2}+2\right)\textup{tan}^2\theta}{\omega^2+\frac{\Gamma^2}{\hbar^4}(n_{\varepsilon}^{T_1}+n_{\varepsilon}^{T_2}+2)^2},\quad \mathcal{A}_{22}^{11}=\mathcal{A}_{11}^{22}=\frac{\frac{\omega^2\Gamma}{2\hbar^2}\left(n_{\varepsilon}^{T_1}+n_{\varepsilon}^{T_2}+2\right)\textup{tan}^2\theta}{\omega^2+\frac{\Gamma^2}{\hbar^4}(n_{\varepsilon}^{T_1}+n_{\varepsilon}^{T_2}+2)^2}\\[0.22cm]
& \mathcal{A}_{gg}^{22}=\frac{2\Gamma}{\hbar^2}n_{\varepsilon}^{T_2},\quad \mathcal{A}_{22}^{22}=-\frac{2\Gamma}{\hbar^2}\left(n_{\varepsilon}^{T_2}+1\right)-\frac{\frac{\omega^2\Gamma}{2\hbar^2}\left(n_{\varepsilon}^{T_1}+n_{\varepsilon}^{T_2}+2\right)\textup{tan}^2\theta}{\omega^2+\frac{\Gamma^2}{\hbar^4}(n_{\varepsilon}^{T_1}+n_{\varepsilon}^{T_2}+2)^2}
\end{aligned}
\label{B.3}
\end{equation}
\end{widetext}
which together with solving Eq.(8) in the main article, lead to the difference between populations of the two excitations $\rho_{22}-\rho_{11}$
\begin{equation}
\begin{split}
\rho_{22}-\rho_{11}= \frac{n_{\varepsilon}^{T_2}-n_{\varepsilon}^{T_1}}{1+2\bar{n}_{\varepsilon}+3n_{\varepsilon}^{T_1}n_{\varepsilon}^{T_2}+\frac{\frac{\omega^2}{4}(\bar{n}_{\varepsilon}+2)\left(3\bar{n}_{\varepsilon}+2\right)\textup{tan}^2\theta}{\omega^2+\frac{\Gamma^2}{\hbar^4}(\bar{n}_{\varepsilon}+2)^2}}
\end{split}
\label{B.4}
\end{equation}
then substituting into the expression of quantum flux $\mathcal{J}_q=\mathcal{A}_{22}^{11}\rho_{22}-\mathcal{A}_{11}^{22}\rho_{11}$ we obtain the form of flux for bosonic case in Eq.(19). For chemical reactions and charge transport, coupled to fermionic baths, the matrix elements are
\begin{widetext}
\begin{equation}
\begin{aligned}
& \mathcal{M}_{gg}^{gg}=-\frac{2\Gamma}{\hbar^2}\left(n_{\varepsilon}^{\mu_1}+n_{\varepsilon}^{\mu_2}\right),\quad \mathcal{M}_{11}^{gg}=\frac{2\Gamma}{\hbar^2}\left(1-n_{\varepsilon}^{\mu_1}\right),\quad \mathcal{M}_{22}^{gg}=\frac{2\Gamma}{\hbar^2}\left(1-n_{\varepsilon}^{\mu_2}\right),\quad \mathcal{M}_{12}^{gg}=\mathcal{M}_{21}^{gg}=0\\
& \mathcal{M}_{gg}^{11}=\frac{2\Gamma}{\hbar^2}n_{\varepsilon}^{\mu_1},\quad \mathcal{M}_{11}^{11}=-\frac{2\Gamma}{\hbar^2}\left(1-n_{\varepsilon}^{\mu_1}\right),\quad \mathcal{M}_{22}^{11}=\mathcal{M}_{11}^{22}=0,\quad \mathcal{M}_{12}^{11}=-\frac{i\omega}{2}\textup{tan}\theta\\
& \mathcal{M}_{21}^{11}=\frac{i\omega}{2}\textup{tan}\theta,\quad \mathcal{M}_{gg}^{22}=\frac{2\Gamma}{\hbar^2}n_{\varepsilon}^{\mu_2},\quad \mathcal{M}_{22}^{22}=-\frac{2\Gamma}{\hbar^2}\left(1-n_{\varepsilon}^{\mu_2}\right),\quad \mathcal{M}_{12}^{22}=\frac{i\omega}{2}\textup{tan}\theta\\
& \mathcal{M}_{21}^{22}=-\frac{i\omega}{2}\textup{tan}\theta,\quad \mathcal{M}_{gg}^{12}=0,\quad \mathcal{M}_{11}^{12}=-\frac{i\omega}{2}\textup{tan}\theta,\quad \mathcal{M}_{22}^{12}=\frac{i\omega}{2}\textup{tan}\theta\\
& \mathcal{M}_{12}^{12}=i\omega-\frac{\Gamma}{\hbar^2}\left(2-n_{\varepsilon}^{\mu_1}-n_{\varepsilon}^{\mu_2}\right),\quad \mathcal{M}_{21}^{12}=0
\end{aligned}
\label{B.5}
\end{equation}
\end{widetext}
then the matrix elements of $\mathcal{A}$ are of the form
\begin{widetext}
\begin{equation}
\begin{aligned}
& \mathcal{A}_{gg}^{gg}=-\frac{2\Gamma}{\hbar^2}\left(n_{\varepsilon}^{\mu_1}+n_{\varepsilon}^{\mu_2}\right),\quad \mathcal{A}_{11}^{gg}=\frac{2\Gamma}{\hbar^2}\left(1-n_{\varepsilon}^{\mu_1}\right),\quad \mathcal{A}_{22}^{gg}=\frac{2\Gamma}{\hbar^2}\left(1-n_{\varepsilon}^{\mu_2}\right),\quad \mathcal{A}_{gg}^{11}=\frac{2\Gamma}{\hbar^2}n_{\varepsilon}^{\mu_1}\\[0.22cm]
& \mathcal{A}_{11}^{11}=-\frac{2\Gamma}{\hbar^2}\left(1-n_{\varepsilon}^{\mu_1}\right)-\frac{\frac{\omega^2\Gamma}{2\hbar^2}\left(2-n_{\varepsilon}^{\mu_1}-n_{\varepsilon}^{\mu_2}\right)\textup{tan}^2\theta}{\omega^2+\frac{\Gamma^2}{\hbar^4}(2-n_{\varepsilon}^{\mu_1}-n_{\varepsilon}^{\mu_2})^2},\quad \mathcal{A}_{22}^{11}=\mathcal{A}_{11}^{22}=\frac{\frac{\omega^2\Gamma}{2\hbar^2}\left(2-n_{\varepsilon}^{\mu_1}-n_{\varepsilon}^{\mu_2}\right)\textup{tan}^2\theta}{\omega^2+\frac{\Gamma^2}{\hbar^4}(2-n_{\varepsilon}^{\mu_1}-n_{\varepsilon}^{\mu_2})^2}\\[0.22cm]
& \mathcal{A}_{gg}^{22}=\frac{2\Gamma}{\hbar^2}n_{\varepsilon}^{\mu_2},\quad \mathcal{A}_{22}^{22}=-\frac{2\Gamma}{\hbar^2}\left(1-n_{\varepsilon}^{\mu_2}\right)-\frac{\frac{\omega^2\Gamma}{2\hbar^2}\left(2-n_{\varepsilon}^{\mu_1}-n_{\varepsilon}^{\mu_2}\right)\textup{tan}^2\theta}{\omega^2+\frac{\Gamma^2}{\hbar^4}(2-n_{\varepsilon}^{\mu_1}-n_{\varepsilon}^{\mu_2})^2}
\end{aligned}
\label{B.6}
\end{equation}
\end{widetext}
which, with the similar procedure, leads to the $\rho_{22}-\rho_{11}$
\begin{equation}
\begin{split}
\rho_{22}-\rho_{11}= \frac{n_{\varepsilon}^{\mu_2}-n_{\varepsilon}^{\mu_1}}{1-n_{\varepsilon}^{\mu_1}n_{\varepsilon}^{\mu_2}+\frac{\omega^2\left(1-\bar{n}_{\varepsilon}^2/4\right)\textup{tan}^2\theta}{\omega^2+\frac{\Gamma^2}{\hbar^4}\left(2-\bar{n}_{\varepsilon}\right)^2}}
\end{split}
\label{B.7}
\end{equation}
which together with the expression of $\mathcal{A}_{22}^{11}$, give the result for fermionic case in Eq.(10). The comparison between analytical and numerical solutions on quantum flux is shown in Fig.9(a) and 9(b) for bosonic and fermionic baths, respectively.
\begin{figure}
\centering\includegraphics[width=3.46in,height=1.4in]{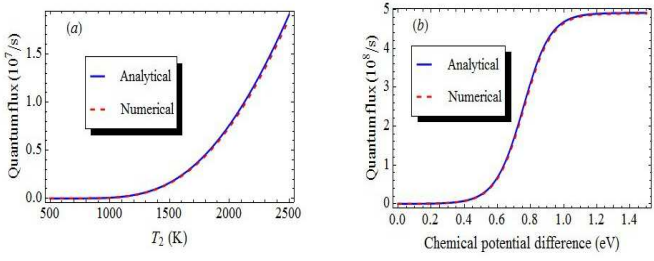}
\caption{(Color online) Analytical and numerical results for quantum flux with (a) bosonic and (b) fermionic reservoirs as a function of bias voltage. Blue(solid) and red(dashed) lines are for analytical and numerical solutions, respectively. Standard parameters are $\varepsilon_1=0.798$eV, $\varepsilon_2=0.8$eV, $\lambda=21$cm$^{-1}$, $\Delta=2$meV and (a) $T_1=500$K, (b) $T=700$K}
\label{Fig.9}
\end{figure}

\section{}
Laplace transform with respect to off-diagonal components of density matrix reads
\begin{equation}
\begin{split}
\tilde{\rho}_{12}=\frac{\mathcal{M}_{gg}^{12}}{s-\mathcal{M}_{12}^{12}} & \tilde{\rho}_{gg}+\frac{\mathcal{M}_{11}^{12}}{s-\mathcal{M}_{12}^{12}}\tilde{\rho}_{11}\\
& +\frac{\mathcal{M}_{22}^{12}}{s-\mathcal{M}_{12}^{12}}\tilde{\rho}_{22}+\frac{\rho_{12}(0)}{s-\mathcal{M}_{12}^{12}}
\end{split}
\label{C.1}
\end{equation}
where $s$ is the Laplace variable in complex frequency domain. It is obvious that the last term controlled by initial conditions in Eq.(\ref{C.1}) vanishes as time approaches $\infty$, after the inverse Laplace transform. Thus a simple initial condition $\rho_{12}(0)=0$ can be applied, as we are only interested in the case of the steady state. The identity of the inverse Laplace transform $\mathcal{F}^{-1}\left(\frac{\tilde{f}(s)}{s-a}\right)=\int_0^{\infty}e^{a(t-u)}f(u)\textup{d}u$ gives the forms of the coherence in terms of populations
\begin{equation}
\begin{split}
\rho_{12} & (t)=\mathcal{M}_{gg}^{12}\int_0^t e^{a(t-t')}\rho_{gg}(t')\textup{d}t'\\
& +\mathcal{M}_{11}^{12}\int_0^t e^{a(t-t')}\rho_{11}(t')\textup{d}t'+\mathcal{M}_{22}^{12}\int_0^t e^{a(t-t')}\rho_{22}(t')\textup{d}t'
\end{split}
\label{C.2}
\end{equation}
By substituting Eq.(\ref{C.2}) into QME we obtain the reduced QME in population space. Of course the dimension of the matrix is reduced from $N^2$ to $N$, the whole system however, becomes memorable, due to the integral over time in the expressions of the coherence components.

The forms of functions $u$ and $v$ in Eq.(19) are
\begin{equation}
\begin{split}
& v^b=\frac{\left(n_{\varepsilon}^{T_2}-n_{\varepsilon}^{T_1}\right)\left(\bar{n}_{\varepsilon}+2\right)}{\left(1+2\bar{n}_{\varepsilon}+3n_{\varepsilon}^{T_1}n_{\varepsilon}^{T_2}\right)\left[1+\frac{\Gamma^2}{\hbar^4\omega^2}\left(\bar{n}_{\varepsilon}+2\right)^2\right]}\\[0.2cm] & u^b=\frac{\left(\bar{n}_{\varepsilon}+2\right)\left(3\bar{n}_{\varepsilon}+2\right)}{4\left(1+2\bar{n}_{\varepsilon}+3n_{\varepsilon}^{T_1}n_{\varepsilon}^{T_2}\right)\left[1+\frac{\Gamma^2}{\hbar^4\omega^2}\left(\bar{n}_{\varepsilon}+2\right)^2\right]}\\[0.2cm]
& v^f=\frac{\left(n_{\varepsilon}^{\mu_2}-n_{\varepsilon}^{\mu_1}\right)\left(2-\bar{n}_{\varepsilon}\right)}{\left[1+\frac{\Gamma^2}{\hbar^4\omega^2}\left(2-\bar{n}_{\varepsilon}\right)^2\right]\left(1-n_{\varepsilon}^{\mu_1}n_{\varepsilon}^{\mu_2}\right)}\\[0.2cm] & u^f=\frac{\left(1-\frac{\bar{n}_{\varepsilon}^2}{4}\right)}{\left[1+\frac{\Gamma^2}{\hbar^4\omega^2}\left(2-\bar{n}_{\varepsilon}\right)^2\right]\left(1-n_{\varepsilon}^{\mu_1}n_{\varepsilon}^{\mu_2}\right)}
\end{split}
\end{equation}
where $\bar{n}_{\varepsilon}\equiv n_{\varepsilon}^{T_1}+n_{\varepsilon}^{T_2}$ or $n_{\varepsilon}^{\mu_1}+n_{\varepsilon}^{\mu_2}$

\textit{Analysis on the flux of quantum systems with non-resonance}-As shown in Fig.2(d) in main article, there is also a sharp increase of flux at vicinity of energy gap $\varepsilon_2$ and then it arrives at saturation, similar to the resonance case. The most significant distinction, however, is the negativity of flux, which indicates that the flux can be reversed although the voltage is positive. This novel behavior is due to the non-resonance within a large gap between excitation energies. In particular, when the chemical potential difference matches the excitation gap $\varepsilon_1$, the site 1 will be much more occupied than site 2 so that the flux is reversed. But in the resonance case, this phenomenon ($\#$ of excitation on site 1$\ll$ $\#$ of excitation on site 2) cannot occur since the two excitation energies are very close to each other, compared to non-resonance case.

\textit{Discussion on temperature effect on flux for the system coupled to chemical reservoirs}-Fig.2(a) shows the temperature effect on the behaviors of quantum flux, as a function of chemical potential difference, for chemical reactions. The curve is step-like function for low temperature while it becomes smooth as the temperature increases, because of the increase of thermal excitations in the vicinity of fermi energy. This character, on the other hand, demonstrates the sharp increase of flux after a critical value of chemical potential $\mu_2$.

\begin{figure}
\centering\includegraphics[width=3.46in,height=1.4in]{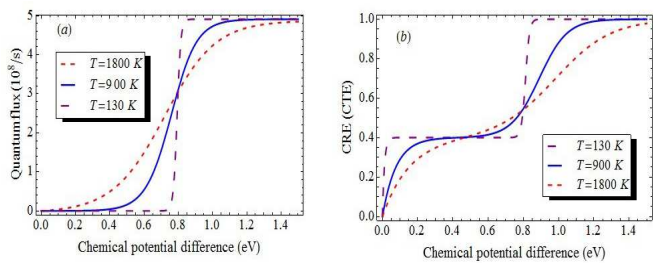}
\caption{(Online Color) Quantum flux and charge transfer efficiency (CRE) vary with respect to voltage, for fermionic reservoirs, with different temperatures. Red, blue and purple lines are for $T=130$K, 900K and 1800K, respectively. Standard parameters are $\varepsilon_1=0.798$eV, $\varepsilon_2=0.8$eV, $\Delta=2$meV and $\mu_1=0$}
\label{Fig.10}
\end{figure}

\section{}
The functions $B$ and $F$ in Eq.(23) of transfer efficiency are of the forms
\begin{equation}
\begin{split}
& B\left(T_1,T_2,\omega\right)=\frac{\left(n_{\varepsilon}^{T_1}+1\right)\left(n_{\varepsilon}^{T_2}+1\right)\left[1+\frac{\Gamma^2}{\hbar^4\omega^2}\left(\bar{n}_{\varepsilon}+2\right)^2\right]}{\bar{n}_{\varepsilon}+2}\\[0.2cm]
& F\left(\mu_1,\mu_2,T,\omega\right)=\frac{\left(1-n_{\varepsilon}^{\mu_1}\right)\left(1-n_{\varepsilon}^{\mu_2}\right)\left[1+\frac{\Gamma^2}{\hbar^4\omega^2}\left(2-\bar{n}_{\varepsilon}\right)^2\right]}{2-\bar{n}_{\varepsilon}}
\end{split}
\end{equation}


\nocite{*}

\begin{thebibliography}{}
\bibitem{Kim05}
Y. Zhang, Y. W. Tan, H. L. Stormer and P. Kim, \textit{Nature} \textbf{438}, 201 (2005)
\bibitem{Laughlin83}
R. B. Laughlin, Phys. Rev. Lett. \textbf{50}, 1395 (1983)
\bibitem{Zhang06}
B. A. Bernevig and S. C. Zhang, Phys. Rev. Lett. \textbf{96}, 106802 (2006)
\bibitem{Hinds05}
M. Majumder, N. Chopra, R. Andrews and B. J. Hinds, \textit{Nature} \textbf{438}, 44 (2005)
\bibitem{Freed72}
W. M. Gelbart, S. A. Rice and K. F. Freed, J. Chem. Phys. \textbf{57}, 4699 (1972)
\bibitem{Kuznetsov99}
A. M. Kuznetsov and J. Ulstrup, \textit{Electron transfer in chemistry and biology: An introduction to the theory}, (Wiley: Chichester, U. K., 1999)
\bibitem{Tour97}
M.\ A.\ Reed, C.\ Zhou, C.\ J.\ Muller, T.\ P.\ Burgin and J.\ M.\ Tour, \textit{Science} \textbf{278}, 252 (1997)
\bibitem{McEuen00}
H.\ Park, J.\ Park, A.\ K.\ L.\ Lim, E.\ H.\ Anderson, A.\ P.\ Alivisatos and P.\ L.\ Mceuen, \textit{Nature} \textbf{407}, 57 (2000)
\bibitem{Umansky05}
M.\ Avinun-Kalish, M.\ Heiblum, O.\ Zarchin, D.\ Mahalu and V.\ Umansky, \textit{Nature} \textbf{436}, 529 (2005)
\bibitem{Mate88}
H.\ Ohtani, R.\ J.\ Wilson, S.\ Chiang and C.\ M.\ Mate, Phys. Rev. Lett. \textbf{60}, 2398 (1988)
\bibitem{Ho04}
S.\ W.\ Wu, G.\ V.\ Nazin, X.\ Chen, X.\ H.\ Qiu and W.\ Ho, Phys. Rev. Lett. \textbf{93}, 236802 (2004)
\bibitem{Ho02}
W.\ Ho, J. Chem. Phys. \textbf{117}, 11033 (2002)
\bibitem{Shluger03}
W.\ A.\ Hofer, A.\ S.\ Foster and A.\ L.\ Shluger, Rev. Mod. Phys. \textbf{75}, 1287 (2003)
\bibitem{Sautet97}
P.\ Sautet, Chem. Rev. \textbf{97}, 1097 (1997)
\bibitem{Esposito06}
U.\ Harbola, M. Esposito and S. Mukamel, Phys. Rev. B \textbf{74}, 235309 (2006)
\bibitem{Skourtis08}
S.\ S.\ Skourtis, D.\ N.\ Beratan, R.\ Naaman, A.\ Niztan and D.\ H.\ Waldeck, Phys. Rev. Lett. \textbf{101}, 238103 (2008)
\bibitem{Schoenmaker02}
B.\ Soree, W.\ Magnus and W.\ Schoenmaker, Phys. Rev. B \textbf{66}, 035318 (2002)
\bibitem{Kubo57}
R. Kubo, J. Phys. Soc. Jap. \textbf{12}, 570 (1957)
\bibitem{Prelovsek05}
D. Baeriswyl and L. Degiorgi, \textit{Strongly Interaction in Low Dimensions} (Springer, 2005)
\bibitem{Saintjam71}
C.\ Caroli, R.\ Combesco, P.\ Norzieres, and D.\ Saintjam, J. Phys. C \textbf{4}, 916 (1971)
\bibitem{Combesco71}
R. Combesco, J. Phys. C \textbf{4}, 2611 (1971)
\bibitem{Car08}
M.\ Koentopp, C.\ Chang, K.\ Burke and R.\ Car, J. Phys.: Condens. Matter \textbf{20}, 083203 (2008)
\bibitem{Mori58}
H.\ Mori, Phys. Rev. \textbf{112}, 1829 (1958)
\bibitem{Baym89}
L.\ P.\ Kadano and G.\ Baym, \textit{Quantum Statistical Mechanics: Green's Function Methods in Equilibrium and Nonequilibrium Problems}  (Addison-Wesley, 1989)
\bibitem{Smith86}
J.\ Rammer and H.\ Smith, Rev. Mod. Phys. \textbf{58}, 323 (1986)
\bibitem{Breuer02}
H. -P.\ Breuer and F.\ Petruccione, \textit{The Theory of Open Quantum Systems} (Oxford University Press, Oxford, 2002)
\bibitem{Spohn80}
H.\ Spohn, Rev. Mod. Phys. \textbf{53}, 569 (1980)
\bibitem{Haake73}
F.\ Haake, \textit{Statistical Treatment of Open Systems}, Springer Tracts in Modern Physics vol. 66 (Springer, Berlin, 1973)
\bibitem{Scully97}
M.\ O.\ Scully and M.\ S.\ Zubairy, \textit{Quantum Optics} (Cambridge University Press, Cambridge, 1997)
\bibitem{Wang12}
L.\ Xu, H.\ Shi, H.\ Feng and J.\ Wang, J. Chem. Phys. \textbf{136}, 165102 (2012)
\bibitem{Qian05}
H.\ Qian, J. Phys.: Condens. Matter \textbf{17}, S3783 (2005)
\bibitem{Wang08}
J.\ Wang, L.\ Xu and E.\ Wang, Proc. Natl. Acad. Sci. U. S. A. \textbf{105}, 12271 (2008)
\bibitem{Wang11}
H.\ Feng and J.\ Wang, J. Chem. Phys. \textbf{135}, 234511 (2011)
\bibitem{Esposito09}
M.\ Esposito and M.\ Galperin, Phys. Rev. B \textbf{79}, 205303 (2009)
\bibitem{Landau77}
L.\ D.\ Landau and E.\ M.\ Lifshitz, \textit{Quantum Mechanics} (\textit{Non-relativistic Theory}), 3rd ed. (Reed Educational and Professional Publishing Ltd., 1977)
\bibitem{Landaustatmech77}
L.\ D.\ Landau, E.\ M.\ Lifshitz and L.\ P.\ Pitaevskij, \textit{Statistical Physics} (\textit{Part 2: Theory of the Condensed State}), 3rd ed. (Reed Educational and Professional Publishing Ltd., 1977)
\bibitem{Qian06}
H. Qian, J. Phys. Chem. B \textbf{110}, 15063 (2006)
\bibitem{Qian09}
H. Qian, Methods Enzymol. \textbf{467}, 111 (2009)
\bibitem{Manzano12}
D.\ Manzano, M.\ Tiersch, A.\ Asadian and H.\ J.\ Briegel, Phys. Rev. E \textbf{86}, 061118 (2012)
\bibitem{Tao08}
F.\ Chen and N.\ J.\ Tao, Acc. Chem. Res. \textbf{42}, 429 (2008)
\bibitem{Sita05}
S.\ A.\ Getty, C.\ Engtrakul, L.\ Wang, R.\ Liu, S.\ H.\ Ke, H.\ U.\ Baranger, W.\ Yang, M.\ S.\ Fuhrer and L.\ R.\ Sita, Phys. Rev. B \textbf{71}, 241401 (2005)
\bibitem{Cao12}
J.\ Ye, K.\ Sun, Y.\ Zhao, Y.\ Yu, C.\ K.\ Lee and J. Cao, J. Chem. Phys. \textbf{136}, 245104 (2012)
\bibitem{Wu12}
J.\ Wu, F.\ Liu, J.\ Ma, R.\ J.\ Silbey and J.\ Cao, J. Chem. Phys. \textbf{137}, 174111 (2012)
\bibitem{Marcus08}
Z.\ Zhu and R.\ A.\ Marcus, J. Chem. Phys. \textbf{129}, 214106 (2008)
\bibitem{Leitner10}
D.\ M.\ Leitner, New J. Phys. \textbf{12}, 085004 (2010)
\bibitem{Wolynes04}
M.\ Gruebele and P.\ G.\ Wolynes, Acc. Chem. Res. \textbf{37}, 261 (2004)
\bibitem{Tao09}
R.\ S.\ Shishir, F.\ Chen, N.\ J.\ Tao and D.\ K.\ Ferry, J. Vac. Sci. Technol. B \textbf{27}, 2003 (2009)
\bibitem{Wolynes92}
Y.\ Tanimura and P.\ G.\ Wolynes, J. Chem. Phys. \textbf{96}, 8485 (1992)
\bibitem{Ghosh11}
P.\ K.\ Ghosh, A.\ Y.\ Smirnov and F.\ Nori, J. Chem. Phys. \textbf{134}, 244103 (2011)
\bibitem{Ghosh09}
P.\ K.\ Ghosh, A.\ Y.\ Smirnov and F.\ Nori, J. Chem. Phys. \textbf{131}, 035102 (2009)
\bibitem{Fleming07}
H.\ Lee, Y.\ C.\ Cheng and G.\ R.\ Fleming, \textit{Science} \textbf{316}, 1462 (2007)
\bibitem{Zhong12}
Y. -T. Kao, X.\ Guo, Y.\ Yang, Z.\ Liu, A.\ Hassanali, Q. -H. Song, L.\ Wang and D.\ Zhong, J. Phys. Chem. B \textbf{116}, 9130 (2012)
\bibitem{Zhong13}
Z.\ Liu, C.\ Tan, X.\ Guo, J.\ Li, L.\ Wang, A.\ Sancar and D. Zhong, Proc. Natl. Acad. Sci. U. S. A. \textbf{110}, 12966 (2013)
\bibitem{Ohmine98}
I.\ Ohmine and S.\ Saito, Acc. Chem. Res. \textbf{32}, 741 (1999)
\bibitem{Xu1992}
D.\ Xu and K.\ Schulten, \textit{The Photosynthetic Bacterial Reaction Center: II. Structure, Spectroscopy and Dynamics}, NATO Science Series A: Life Sciences, 301-312 (Plenum Press, New York, 1992)
\bibitem{Wolynes1988}
J.\ N.\ Onuchic and P.\ G.\ Wolynes, J. Phys. Chem. \textbf{92}, 6495 (1988)
\bibitem{Qian04}
M.\ P.\ Qian and M.\ Qian, Zeitschrift fur Wahrscheinlichkeitstheorie
und Verwandte Gebiete. \textbf{59(2)}, 203(1982); M.\ Qian and Z.\ T.\ Hou, \textit{Reversible Markov Process}, (Hunan Scientific Publisher, Changsha, 1979).
\bibitem{Zia}    
R.\ K.\ P. Zia and B.\ Schmittmann, J. of Stat. Mech.: Theory and Exp. \textbf{7(7)}, 07012 (2007).
\bibitem{collini10}
E.\ Collini, C.\ Y.\ Wong, K.\ E.\ Wilk, P.\ M.\ G.\ Curmi, P.\ Brumer and G.\ D.\ Scholes, \textit{Nature} \textbf{463}, 644 (2010)
\bibitem{JPCL2013}
I.\ Kassal, J.\ Yuen-Zhou and S.\ R.\ Keshari, J. Phys. Chem. Lett., \textbf{4}, 362 (2013)
\bibitem{inter2014}
F.\ Fassioli, R.\ Dinshaw, P.\ C.\ Arpin and G.\ D.\ Scholes, \textit{J. R. Soc. Interface}, \textbf{11}, 20130901 (2014)
\end{thebibliography}

\end{document}